\documentclass[preprint,superscriptaddress]{revtex4}
\usepackage{graphicx}
\usepackage{array}
\usepackage{amssymb}
\usepackage{amsfonts}
\usepackage{amsmath}
\usepackage{mathrsfs}
\usepackage{color}
\usepackage{booktabs}
\usepackage{threeparttable}
\usepackage{multirow}
\usepackage{subfigure}
\usepackage{times}
\usepackage{epsfig}
\usepackage{threeparttable}
\usepackage{chngpage}
\usepackage{latexsym}

\begin{document}

\title{Control and controllability of nonlinear dynamical networks: a 
geometrical approach}
\date\today

\author{Le-Zhi Wang}
\affiliation{School of Electrical, Computer and Energy Engineering, Arizona State University, Tempe, Arizona 85287, USA}

\author{Ri-Qi Su}
\affiliation{School of Electrical, Computer and Energy Engineering, Arizona State University, Tempe, Arizona 85287, USA}

\author{Zi-Gang Huang}
\affiliation{School of Electrical, Computer and Energy Engineering, Arizona State University, Tempe, Arizona 85287, USA}
\affiliation{Institute of Computational Physics and Complex Systems, Lanzhou University, Lanzhou Gansu 730000, China}

\author{Xiao Wang}
\affiliation{School of Biological and Health Systems Engineering, Arizona State University, Tempe, AZ 85287, USA}

\author{Wenxu Wang}
\affiliation{School of Systems Science, Beijing Normal University,
Beijing, 100875, P. R. China}
\affiliation{School of Electrical, Computer and Energy Engineering, Arizona State University, Tempe, Arizona 85287, USA}

\author{Celso Grebogi}
\affiliation{Institute for Complex Systems and Mathematical Biology,
King's College, University of Aberdeen, Aberdeen AB24 3UE, UK}

\author{Ying-Cheng Lai} \email{Ying-Cheng.Lai@asu.edu}
\affiliation{School of Electrical, Computer and Energy Engineering, Arizona State University, Tempe, Arizona 85287, USA}
\affiliation{Institute for Complex Systems and Mathematical Biology,
King's College, University of Aberdeen, Aberdeen AB24 3UE, UK}
\affiliation{Department of Physics, Arizona State University, Tempe, Arizona 85287, USA}

\begin{abstract}

{\bf In spite of the recent interest and advances in linear 
controllability of complex networks, controlling nonlinear network
dynamics remains to be an outstanding problem.
we develop an experimentally feasible control framework for nonlinear 
dynamical networks that exhibit multistability (multiple coexisting
final states or attractors), which are representative of, e.g., gene 
regulatory networks (GRNs). The control objective is to apply parameter 
perturbation to drive the system from one attractor to another, assuming 
that the former is undesired and the latter is desired. To make our
framework practically useful, we consider {\em restricted} parameter
perturbation by imposing the following two constraints: (a) it must 
be experimentally realizable and (b) it is applied only temporarily.  
We introduce the concept of {\em attractor network}, in which
the nodes are the distinct attractors of the system, and 
there is a directional link from one attractor to another if the system
can be driven from the former to the latter using restricted control
perturbation. Introduction of the attractor network allows us to formulate 
a controllability framework for nonlinear dynamical networks: a 
network is more controllable if the underlying attractor network 
is more strongly connected, which can be quantified. We demonstrate our 
control framework using examples from various models of experimental
GRNs. A finding is that, due to nonlinearity, noise can 
counter-intuitively facilitate control of the network dynamics.}
 
\end{abstract}
\maketitle

An outstanding problem in interdisciplinary research is
to control nonlinear dynamics on complex networks. Indeed, the physical 
world in which we live is nonlinear, and complex networks are ubiquitous 
in a variety of natural, social, economical, and man-made systems. Dynamical 
processes on complex networks are thus expected to be generically 
nonlinear. While the ultimate goal to study complex systems is to control 
them, the coupling between nonlinear dynamics and complex network 
structures presents tremendous challenges to our ability to formulate 
effective control methodologies. In spite of the rapid development of 
network science and engineering toward understanding, analyzing and 
predicting the dynamics of large complex network systems in the past 
fifteen years, the problem of controlling nonlinear dynamical networks has 
remained open.

In the past several years, a framework for determining the \emph{linear} 
controllability of network based on traditional control and graph theories 
emerged~\cite{LH:2007,LCWX:2008,RJME:2009,LSB:2011,WNLG:2011, YZDWL:2013,
CWWL:2015,NA:2012,YRLLL:2012,NV:2012,LSBA:2013,MDB:2014,RR:2014,Wuchty:2014,
WBSS:2015,YTBSLB:2015}, leading to a quantitative understanding of the 
effect of network structure on its controllability. In particular, a 
structural-controllability framework was proposed~\cite{LSB:2011}, revealing 
that the ability to steer a complex network toward any desired state, as 
measured by the minimum number of driver nodes, is determined by the set 
of maximum matching, which is the maximum set of links that do not share 
starting or ending nodes. A main result was that the number of driver 
nodes required for full control is determined by the network's degree 
distribution~\cite{LSB:2011}. The framework was established for weighted 
and directed networks. An alternative framework, the exact-controllability 
framework, was subsequently formulated~\cite{YZDWL:2013}, which 
was based on the principle of maximum multiplicity to identify the 
minimum set of driver nodes required to achieve full control of networks 
with arbitrary structures and link-weight distributions. Generally, the 
deficiency of such rigorous mathematical frameworks of controllability is 
that the nodal dynamical processes must be assumed to be \emph{linear}. For 
nonlinear nodal dynamics, the mathematical framework on which the 
controllability theories based, namely the classic Kalman's controllability 
rank condition~\cite{Kalman:1963,Lin:1974,Luenberger:book}, is not 
applicable. At the present there is no known theoretical framework for 
controlling nonlinear dynamics on complex networks. 

Due to the high dimensionality of nonlinear dynamical networks and the 
rich variety of behaviors that they can exhibit, it may be 
prohibitively difficult to develop a control framework that is
universally applicable to different kinds of network dynamics. 
In particular, the classic definition of linear controllability,
i.e., a network system is controllable if it can be driven from an arbitrary
initial state to an arbitrary final state in finite time, is generally
not applicable to nonlinear dynamical networks. Instead,
controlling collective dynamical behaviors may be more pertinent and
realistic~\cite{WC:2002,LWC:2004,SBGC:2007,CZL:2014}. 
Our viewpoint is that, for nonlinear 
dynamical networks, control strategies may need to be specific and system 
dependent. The purpose of this paper is to articulate control strategies
and develop controllability framework for nonlinear networks that exhibit 
multistability. A defining characteristic of such 
systems is that, for a realistic parameter setting, there are multiple
coexisting attractors in the phase 
space~\cite{GMOY:1983,MGOY:1985,FG:1997,FG:2003,LT:book,NYLDG:2013}.
The goal is to drive the system
from one attractor to another using physically meaningful, temporary and
finite parameter perturbations, assuming that the system is likely to 
evolve into an undesired state (attractor) or the system is already in 
such a state, and one wishes to implement control to bring the system out 
of the undesired state and steer it into a desired one. We note that
dynamical systems with multistability are ubiquitous in the real world
ranging from biological and ecological to physical 
systems~\cite{Alley:2003,May:1977,SPD:2005,Chase:2003,BM:2005,WZXW:2011,
HGME:2007}. 

In biology, nonlinear dynamical networks with multiple attractors have 
been employed to understand fundamental phenomena such as cancer
mechanisms~\cite{Huang:2013}, cell fate differentiation~\cite{SGLE:2006,
HEYI:2005,FK:2012,LZW:2014}, and cell cycle control~\cite{YTWNJY:2011,
BT:2004}. For example, boolean network models were used to study the 
gene evolution~\cite{KPST:2004}, attractor number variation with 
asynchronous stochastic updating~\cite{GD:2005}, gene expression in 
the state space~\cite{HEYI:2005}, and organism system growth rate 
improvement~\cite{MGAB:2008}. Another approach is to abstract key regulation 
genetic networks~\cite{MTELT:2009,Faucon:2014} (or motifs) from all 
associated interactions, and to employ synthetic biology to modify, 
control and finally understand the biological mechanisms within these 
complicated systems~\cite{YTWNJY:2011,SGLE:2006}. An earlier application
of this approach led to a good understanding of the ubiquitous phenomenon
of bistability in biological systems~\cite{Gardner:2000}, where there are
typically limit cycle attractors and, during cell cycle control, noise can
trigger a differentiation process by driving the system from a limit circle 
to another steady state attractor~\cite{SGLE:2006}. Generally speaking,  
there are two candidate mechanisms for transition or switching between 
different attractors~\cite{FK:2012}: through signals transmitted between
cells and through noise, which were demonstrated recently using synthetic 
genetic circuits~\cite{WSLELW:2013,WMX:2014}. More recently, a detailed 
numerical study was carried out of how signal-induced bifurcations in a 
tri-stable genetic circuit can lead to transitions among different cell 
types~\cite{LZW:2014}.

In this paper, we develop a controllability framework for nonlinear 
dynamical networks based on the concept of {\em attractor 
networks}~\cite{Lai:2014}. An attractor network is defined in the phase 
space of the underlying nonlinear system, in which each node represents an 
attractor and a directed edge from one node to another indicates that 
the system can be driven from the former to the latter using experimentally
feasible, temporary, and finite parameter changes. A well connected 
attractor network implies a strong feasibility that the system can be
controlled to reach a desired attractor. The connectivity of the 
attractor network can then be used to characterize the controllability 
of the nonlinear network. More specifically, for a given pair of attractors,
the weighted shortest path between them in the attractor network is an
indicator of the physical feasibility of the associated transition. We 
use gene regulatory networks (GRNs) to demonstrate our control framework, 
which includes low-dimensional, experimentally realizable synthetic 
gene circuits and a realistic T-cell cancer network of 60 nodes. A 
finding is that noise can counter-intuitively enhance the 
controllability of a nonlinear dynamical network. We emphasize that the
development of our nonlinear control framework is based entirely on 
physical considerations, rendering feasible experimental verification. \\
\\ \noindent
{\large\bf Results} \\
\\ \noindent
A complex, nonlinear dynamical network of $N$ variables can be described 
by a set of $N$ coupled differential equations: 
\begin{equation} \label{eq:sys}
\dot{\mathbf{x}}=\mathbf{F}(\mathbf{x},{\bf \mu}),
\end{equation}
where $\mathbf{x}\in \mathbf{R}^N$ denotes the $N$-dimensional state 
variable, $\mathbf{F}(\mathbf{x},{\bf \mu})$ is the nonlinear vector field, 
and ${\bf \mu} \in \mathbf{R}^M$ represents the set of coupling parameters.
In a GRN, the nodal dynamics is typically one dimensional. For simplicity, 
we assume that this is the case to be treated so that the size 
of the network represented by Eq.~(\ref{eq:sys}) is $N$. From 
consideration of realistic GRNs, we assume that the coupling 
parameters can be adjusted externally, which are effectively the set of 
{\em control parameters}. Specifically, in a GRN, the various coupling 
strengths among the nodes (genes) can be experimentally and systematically 
varied through application of specific targeted drugs. At a larger scale, 
the fate of a cell can be controlled by adding drugs to the 
cell-growth environment, which adjust the interaction parameters in the 
underlying network~\cite{Gardner:2000}. While dynamical variables 
themselves can also be perturbed for the purpose of control, for GRNs 
this is unrealistic. For this reason the scenario of perturbing dynamical 
variables will not be considered in this paper. 

\begin{figure*}
\begin{center}
\epsfig{file=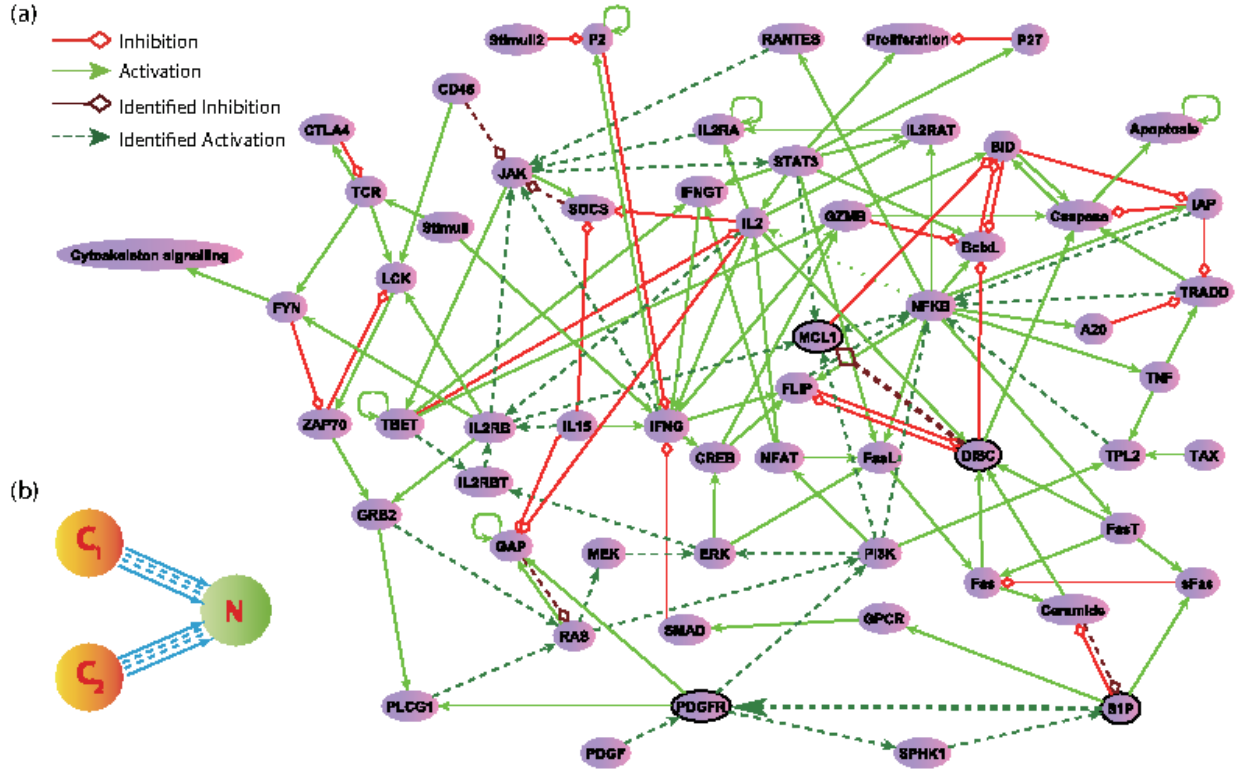,width=1.0\linewidth}
\caption{\small {\bf T-LGL survival signaling system and its attractor 
network}. (a) Structure of T-LGL signaling network: each node is 
labeled with its generic name, and the arrowhead and diamond-head edges 
represent excitation and inhibition regulations, respectively. The 
inhibitory edges from ``Apoptosis'' to other nodes are not shown
(for clarity). (b) Attractor network of the T-cell network, which 
contains three nodes: two cancerous states denoted as $\mathbf C_1$ and 
$\mathbf C_2$ and a normal state denoted as $\mathbf N$. Our detailed 
computations reveal that parameter perturbations on $48$ edges can drive 
the system from a cancerous state to the normal state, which are indicated 
with dark dashed lines, whereas the remaining edges in the network are 
specified with light solid lines. The two directed edges in the attractor 
network are multiple, containing altogether $48$ individual edges 
corresponding to controlling the $48$ solid-line edges in the 
original network.}
\label{fig:structTcell}
\end{center}
\end{figure*}

We focus on nonlinear dynamical networks with \emph{multiple} coexisting
attractors. For a given set of parameters ${\bf \mu}$, the multiple 
attractors (e.g., stable steady states) and the corresponding basins are 
fixed. For a given initial condition, the system will approach one of 
the attractors. Each attractor has specific biological significance, which
can be regarded as either desired or undesired, depending on the particular
function of interest. Suppose, without any control, the system is in  
an undesired attractor or is in its basin of attraction. The question is 
how to steer the system from the undesired state to a desired state by means 
of {\em temporal and small} parameter variations that are experimentally 
feasible. 

\paragraph*{Control principle based on bifurcation.}
To motivate the development of a feasible control principle, we consider 
the simple case where the system is near a bifurcation and control is to be 
applied to drive the system from one attractor to another through temporal 
perturbation to a \emph{single} parameter. That is, the parameter variation
is turned on and takes effect for a finite (typically short) duration of
time. After control perturbation is withdrawn, the system is restored to
its parameter setting before control but its state has been changed: it 
will be in the desired attractor. Let $\mu_0$ be the initial 
parameter value and the system is in an undesired attractor denoted
as $\mathbf{x}_i^\ast$, and let $\mathbf {x}_f^\ast$ be the desired 
attractor that the system is driven to. Imposing control means that we
change the parameter from $\mu_0$ to $\mu_1$. The dynamical mechanism 
to drive the system out of the initial attractor is bifurcations, e.g., 
a saddle-node bifurcation at which the original attractor disappears and 
its basin is absorbed into that of an \emph{intermediate 
attractor}~\cite{BT:2004}, denoted as ${\mathbf{\bar x}}_k^\ast$. Turning
on control to change $\mu$ from $\mu_0$ to $\mu_1$ thus makes the system
approach ${\mathbf{\bar x}}_k^\ast$. This process continues until the 
system falls into the original basin of $\mathbf {x}_f^\ast$, at which 
point the control parameter is reset to its original value $\mu_0$ so 
that the system will approach the desired attractor $\mathbf{x}_f^\ast$. 
Success of control relies on the existence of a ``path'' from the initial
attractor to the final one through a number of intermediate attractors. 
If a single parameter is unable to establish such a path, variations in 
multiple parameters can be considered, {\em provided that such parameter 
adjustments are experimentally realizable}. For a biological network, 
this can be achieved through application of a combined set of 
drugs~\cite{Feala:2010,Fitzgerald:2006}. However, even when potential 
complications induced by inter-drug interactions are neglected, the 
search space for suitable parameter perturbation can be prohibitively 
large if we allow all available parameters to be adjusted simultaneously. 
We demonstrate below that this challenge can be met by constructing 
an \emph{attractor network} for the underlying system.

\paragraph*{Attractor networks: an example of T-cell network}
For a complex, nonlinear dynamical network, an attractor network can 
be constructed by defining each of all possible attractors of the system 
as a node. There exists a directed link from one node to another if an
experimentally accessible parameter of the system can be adjusted to 
drive or control the system from the former to the latter. There can be 
multiple edges from one node to another, if there are multiple parameters, 
each enabling control. Starting from an initial attractor, one can identify,
using all accessible parameters with variations in physically reasonable
ranges, a set of attractors that the system can be driven into. Repeating
this procedure for all attractors in the system, we build up an attractor
network that provides a {\em blueprint} for driving the whole networked system
from any attractor to any other attractor in the system, assuming at the
time the latter attractor would lead to desired function of the system
as a whole. All these can be done using relatively small parameter 
perturbations.

To demonstrate the construction of an attractor network, we take as an
example a realistic biological network, T-cell in large granular 
lymphocyte leukemia associated with blood cancer. Specifically, apoptosis 
signaling of the T-cell can be described by a network model: T-cell survival 
signaling network~\cite{Zhang:2008,Saadatpour:2011}, which has 60 nodes 
and 142 regulatory edges, as shown in Fig.~\ref{fig:structTcell}(a). 
Nodes in the network represent proteins and transcripts, and the edges 
correspond to either \emph{activation} or \emph{inhibitory} regulations. 
Experimentally, it was found that there are three attractors for this 
biophysically detailed network~\cite{Zhang:2008,Saadatpour:2011}. 
Among the three attractors, two correspond to two distinct cancerous 
states (denoted as $\mathbf C_1$ and $\mathbf C_2$) and one is associated 
with a normal state (denoted as $\mathbf N$). 
By translating the Boolean rules into a continuous form using the method 
in~\cite{Wittmann:2009,KPWT:2010} and setting the strength of each edge to 
unity, one can obtain a set of nonlinear dynamical equations for the entire
network system. Direct simulation of the model revealed that there are 
three stable fixed point attractors, in agreement with the experimental
observation~\cite{Zhang:2008,Saadatpour:2011}. 
The attractor network is thus quite simple: it has three
nodes only, as shown in Fig.~\ref{fig:structTcell}(b). Testing all 
experimentally adjustable parameters, we find multiple edges from each 
cancerous attractor to the normal one (see Supplementary Table 1). Since
the goal of control is to bring the system from one of the cancerous states
to the normal one, it is not necessary (or biologically meaningful) to test 
whether parameter perturbation exists that drives the system from the 
normal node to a cancerous node.  

\begin{figure}
\begin{center}
\epsfig{figure=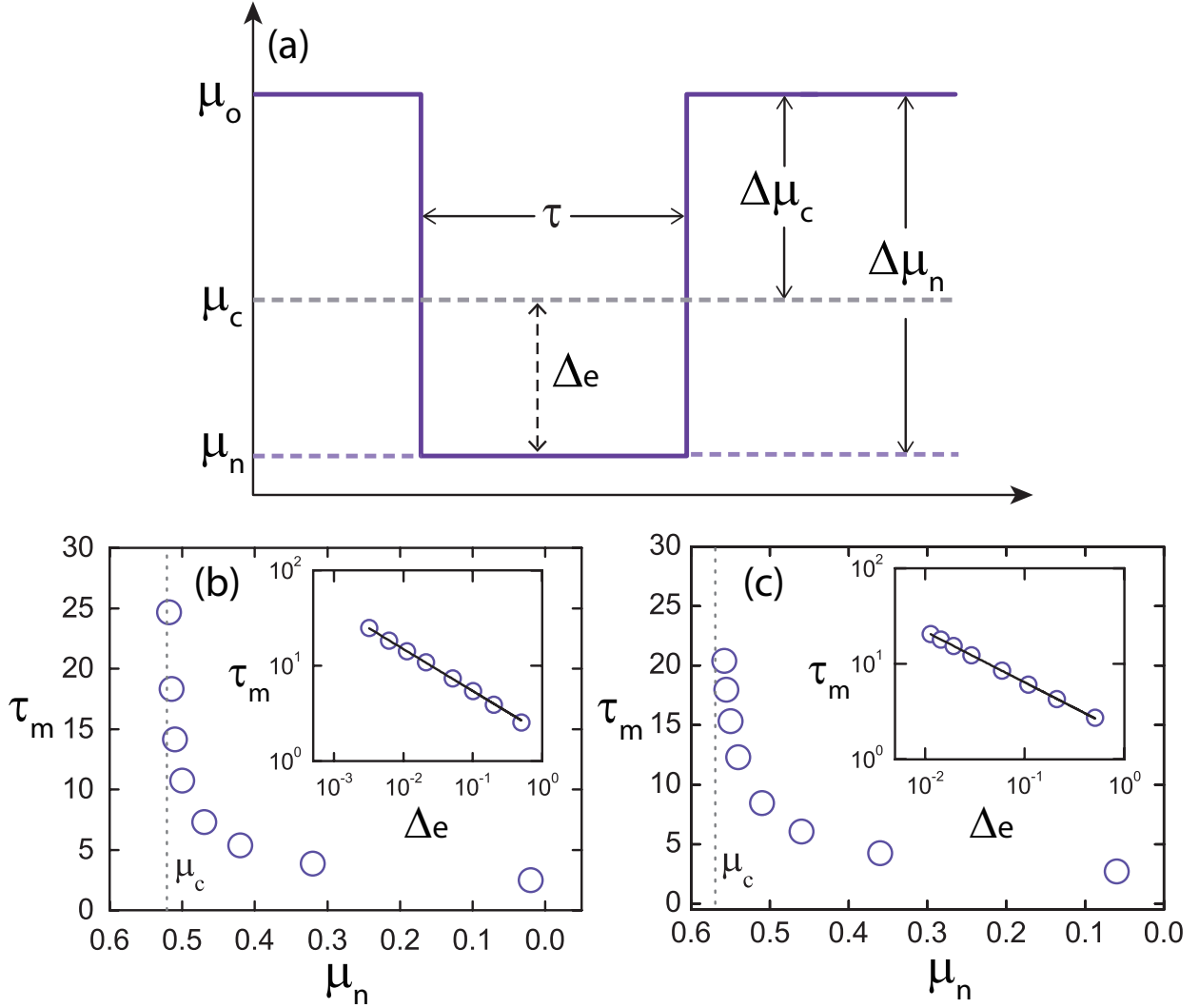,width=\linewidth}
\caption{\small {\bf Relationship between edge control strength and minimal 
control time}. For the T-cell network, (a) an inverted rectangular control 
signal of duration $\tau$ and amplitude $\Delta \mu=|\mu_n-\mu_0|$, where 
$\mu_0$ is the original parameter value. A saddle-node bifurcation occurs 
for $\mu = \mu_c$, so $\Delta e = \mu_c - \mu_n$ is the excess amount of 
the parameter change over the critical value $\mu_c$. (b,c) Minimal 
control time $\tau_m$ versus $\mu_n$, where parameter control is applied 
to the activation edge from node ``S1P'' to node ``PDGFR'' and to the 
inhibitory edge from ``DISC'' to ``MCL1'', respectively. These four nodes 
are indicated using solid black circles in Fig.~\ref{fig:structTcell}(a). The 
corresponding plots on a logarithmic scale in the insets of (b) and (c) 
suggest a power-law scaling behavior between $\tau_m$ and $\Delta e$
[Eq.~(\ref{eq:taum})]. The fitted power-law scaling exponents are 
$\beta \approx 0.42$, and $0.38$, respectively, for (b) and (c).}
\label{fig:uc_tmtcell}
\end{center}
\end{figure}

\paragraph*{Control implementation based on attractor network.}
Given a nonlinear dynamical network in the real (physical) space, the
underlying phase space dimension may be quite high, rendering analysis of 
the dynamical behaviors difficult. The attractor network is a coarse grained
representation of the phase space, retaining information that is most
relevant to the control task of driving the network system to a desired
final state. Once an attractor network has been constructed, actual 
control can be carried out through temporary changes in a set of 
experimentally adjustable parameters, one at a time. This should be 
contrasted to one existing approach~\cite{CKM:2013} that requires accurate 
adjustments in the state variables, which may not always be realistic. 

We detail how actual control can be implemented based on the attractor
network for the T-cell network. To be concrete, we assume that the control
signal has the shape of a rectangular pulse in the plot of a parameter 
versus the time, as shown in Fig.~\ref{fig:uc_tmtcell}(a), where the control
parameter is $\mu$ and the rectangular pulse has duration $\tau$ and 
amplitude $\Delta \mu=|\mu_n-\mu_0|$, with $\mu_0$ denoting the nominal 
parameter value and $\mu_n$ being the value during the time interval when
control is on. For the T-cell network, we set $\mu_0 = 1.0$. As $\mu$ is 
reduced the system approaches a bifurcation point. (In other examples
a bifurcation can be reached by increasing a control parameter, as in 
low-dimensional GRNs detailed in {\bf Control Analysis}.) Extensive 
numerical simulations in controlling the T-cell network from a cancerous
state ($\mathbf C_1$ or $\mathbf C_2$) to the normal state $\mathbf N$ 
shows that, to achieve control, there are wide ranges of choices for 
$\Delta \mu$ and $\tau$. In fact, once $\mu_n$ is decreased through the 
bifurcation point $\mu_c$ at which the initial attractor loses its 
stability, the goal of control can be realized. The critical value 
$\mu_c$ for each parameter can be identified from a bifurcation analysis. 
Additionally, for $\mu_n < \mu_c$, there exists a required minimum control 
time $\tau_m$, over which the system will move into the original basin of the 
target attractor before control is activated. Insofar as $\tau > \tau_m$, the 
control signal can be released. Longer duration of control is not necessary 
since the system will evolve into the target attractor following its natural 
dynamical evolution associated with the nominal parameter $\mu_0$. The value 
of $\tau_m$ increases as $\mu_n$ is closer to $\mu_c$, where if 
$\mu_n=\mu_c$, $\tau_m$ is infinite due to the critical slowing down 
at the bifurcation point $\mu_c$. Figures~\ref{fig:uc_tmtcell}(b) and
\ref{fig:uc_tmtcell}(c) show, respectively, for the T-cell network, the 
relationship between $\tau_m$ and $\mu_n$ in controlling the strength of 
the activation edge from node ``S1P'' to node ``PDGFR'', and that of the 
inhibitory edge from node ``DISC'' to node ``MCL1'' 
[cf., Fig.~\ref{fig:structTcell}(a), the 
nodes denoted as black circles and connected by bold coupling edges]. 
The critical value $\mu_c$ (indicated by the dotted line) can be estimated 
accordingly. The insets of (b) and (c) show the corresponding plots of 
the relationships on a double logarithmic scale, with the horizontal 
axis to be $\Delta_\mathrm{e}=\mu_c-\mu_n$, the \emph{exceeded} value 
of $\mu_n$ over the critical point $\mu_c$. We observe the following
power-law scaling behavior:
\begin{equation} \label{eq:taum}
\tau_m = \alpha |\mu_n-\mu_c |^{\beta},
\end{equation}
where $\beta$ is the scaling exponent. The region of control parameters at 
the upper-right region over the curve of $\tau_m(\Delta e)$, i.e., larger 
$\Delta e$ value or longer duration $\tau$, corresponds to the case where
control is successful in the sense that the system can definitely be 
driven to the desired final state.

The power-law scaling relation for $\tau_m$ demonstrated in 
Figs.~\ref{fig:uc_tmtcell}(b) and~\ref{fig:uc_tmtcell}(c) for the T-cell
network is quite general, as it also holds for two-node and     
three-node GRNs (see {\bf Control Analysis}). For the T-cell system, the 
critical values of parameters for all the possible controllable edges 
from $\mathbf C_1$ or $\mathbf C_2$ to $\mathbf N$, and the corresponding 
values of $\alpha$ and $\beta$ in Eq.~(\ref{eq:taum}) are provided in
Table~S1 in Supplemental Information). The control magnitude 
and time for some parameters are identical, for the reason that the logic 
relationship from the corresponding edges to the same node can be 
described as ``AND'' (c.f., Fig.~\ref{fig:structTcell}) so that in the
continuous-time differential equation model, all these in-edges are 
equivalent. (The control results from the two-node and three-node GRNs
between any pair of nearest-neighbor attractors are listed in Tables~S2 
and S3 in \textbf{SI}, respectively.) Due to the flexibility in choosing
the control signal, our control scheme based on the attractor network is 
amenable to experimental implementation.

\begin{figure}
\begin{center}
\epsfig{figure=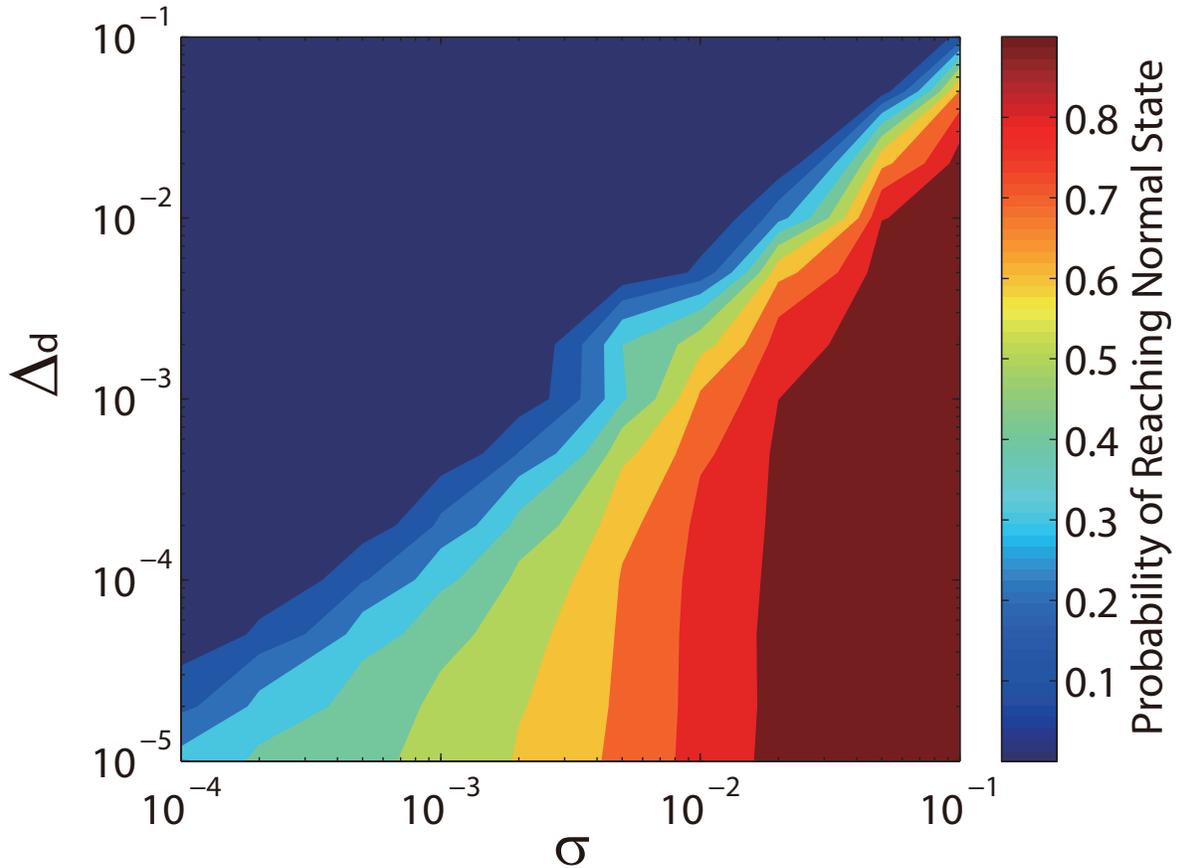,width=\linewidth}
\caption{\small {\bf Beneficial role of noise in controlling the T-cell
network: probability that the normal state can be reached}. Success rate 
to control the T-cell system from the cancerous state $\mathbf C_1$ to the 
normal state $\mathbf N$ using a combination of parameter perturbation 
and external noise (of amplitude) $\sigma$, where 
$\Delta_\mathrm{d} \ \mu_n - \mu_c$ is the parameter deficiency. Warm colors 
indicate higher probability values of successful control. The perturbation
duration is $\tau=200$. The results are averaged over $1000$ realizations.}
\label{fig:uc_noiseTcell}
\end{center}
\end{figure}

\paragraph*{Beneficial role of noise in control.}
More than three decades of intense research in nonlinear dynamical systems 
has led to great knowledge about the role of noise, in terms of phenomena 
such as stochastic 
resonance~\cite{BSV:1981,BPSV:1983,MW:1989,MPO:1994,GNCM:1997,GHJM:1998}, 
coherence resonance~\cite{SH:1989,PK:1997,LL:2001a,LL:2001b}, 
and noise-induced chaos~\cite{LT:book}, etc. Common to all these phenomena 
is that a proper amount of noise can in fact be beneficial, for example, 
to optimize the signal-to-noise ratio, to enhance the signal regularity 
or temporal coherence, or to facilitate the transitions among the attractors.
We find that, in our attractor-network based control framework, noise can
also be beneficial. This can be understood intuitively by 
noting that our control mechanism is to make the system leave an undesired
attractor and approach a desired one but noise in combination with 
parameter adjustments can facilitate the process of escaping from 
an attractor. To demonstrate this, we assume that 
the T-Cell network is subject to Gaussian noise, which can be modeled by 
adding independent normal distribution terms $\mathbf{N}(0,\sigma^2)$ to 
the system equations, where $\sigma$ is the noise amplitude. We find that,
with noise, the required magnitude of parameter change can be reduced. 
In fact, even when the controlled parameter $\mu_n$ has not yet reached 
the bifurcation point $\mu_c$, noise can lead to a non-zero probability 
for the system to escape the basin of the undesired attractor. 

Suppose the control parameter is set to the value $\mu_n$, which is 
insufficient to induce escape from the undesired attractor without noise. 
When noise is present, the system dynamics is stochastic. 
To characterize the control performance, 
we use a large number of independent realizations with the same 
initial condition. Specifically, we perform independent simulations 
starting from one cancerous state, e.g., $\mathbf C_1$, but with insufficient 
control strength as characterized by the deficiency parameter 
$\Delta_\mathrm{d} \equiv \mu_n - \mu_c$, and calculate the probability $P$
of control success through the number of trials that the system can be
successfully driven to the normal state $\mathbf N$.
Figure~\ref{fig:uc_noiseTcell} shows, on a double logarithmic scale, the 
values of $P$ in the parameter plane of $\sigma$ and 
$\Delta_\mathrm{d}$, where the control parameter is
the strength of the activation edge from node ``S1P'' to node ``PDGFR'' in 
the T-cell network. We see that, for fixed $\sigma$, $P$ decreases with
$\Delta_\mathrm{d}$ but, for any fixed value of $\Delta_\mathrm{d}$, 
the probability $P$ increases with $\sigma$, indicating the beneficial role
of noise in facilitating control. In the parameter plane there exists a 
well-defined boundary, below which the control probability assumes large 
values but above which the probability is near zero. Testing alternative 
control parameters yields essentially the same results, due to the 
simplicity of the attractor network for the T-cell system and the 
multiple directed edges from each cancerous state to the normal state. \\
\\ \noindent
{\large\bf Control Analysis} \\
\\ \noindent
In spite of the simplicity of its attractor network, the original T-cell 
network itself is still quite complicated from the point of view of 
nonlinear dynamical analysis. To have a better understanding of our control
mechanism, we study GRNs of relatively low dimensions and carry out a 
detailed analysis of the associated attractor networks.   

\begin{figure}
\begin{center}
\epsfig{file=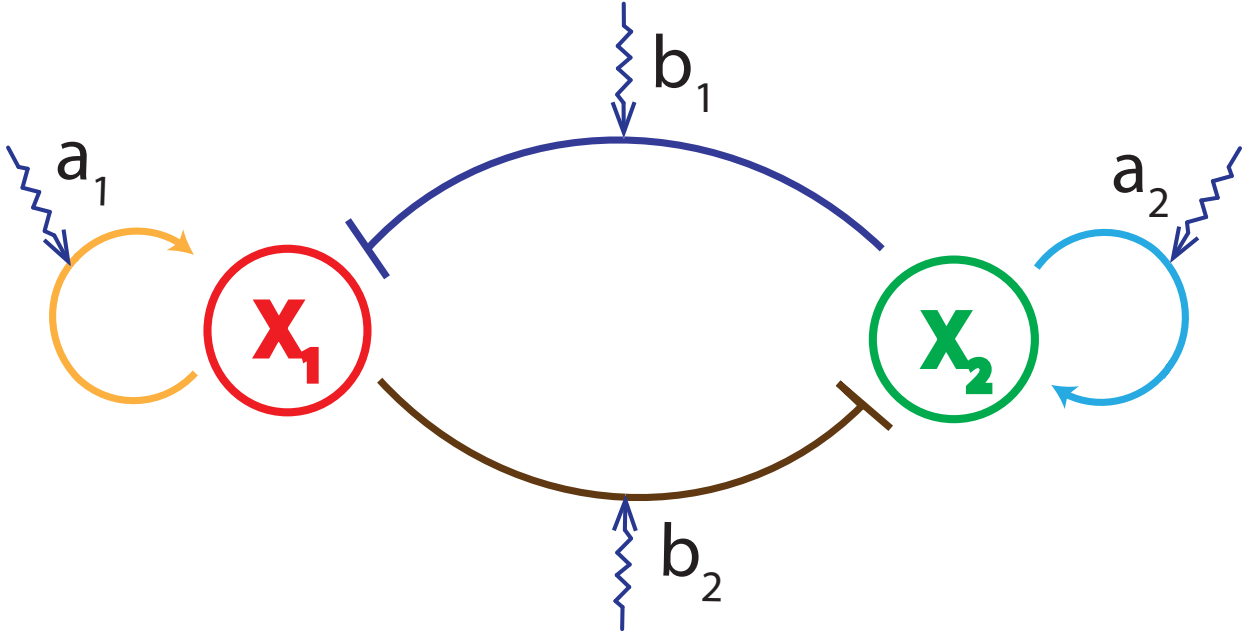,width=\linewidth}
\caption{\small {\bf A two-node GRN.} 
Simplified model of the two-node GRN, where the arrowhead and bar-head 
edges represent activation and inhibition regulation, respectively. The 
sawtooth lines denote the strength of the tunable edge.} 
\label{fig:top2D}
\end{center}
\end{figure}

\paragraph*{Attractor network for a two-node GRN.}
We use a two-node GRN to understand the dynamical mechanism underlying 
the attractor network. As shown in Fig.~\ref{fig:top2D}, the network  
contains two auto-activation nodes (genes) and together they form a 
mutual inhibitory circuit. Such a topology was shown to be responsible 
for the regulation of blood stem cell differentiation~\cite{HGME:2007}. 
In addition, it is conceivable that such topologies can be constructed 
with tunable inputs using synthetic biology approaches~\cite{WSLELW:2013}.

\begin{figure*}
\begin{center}
\epsfig{file=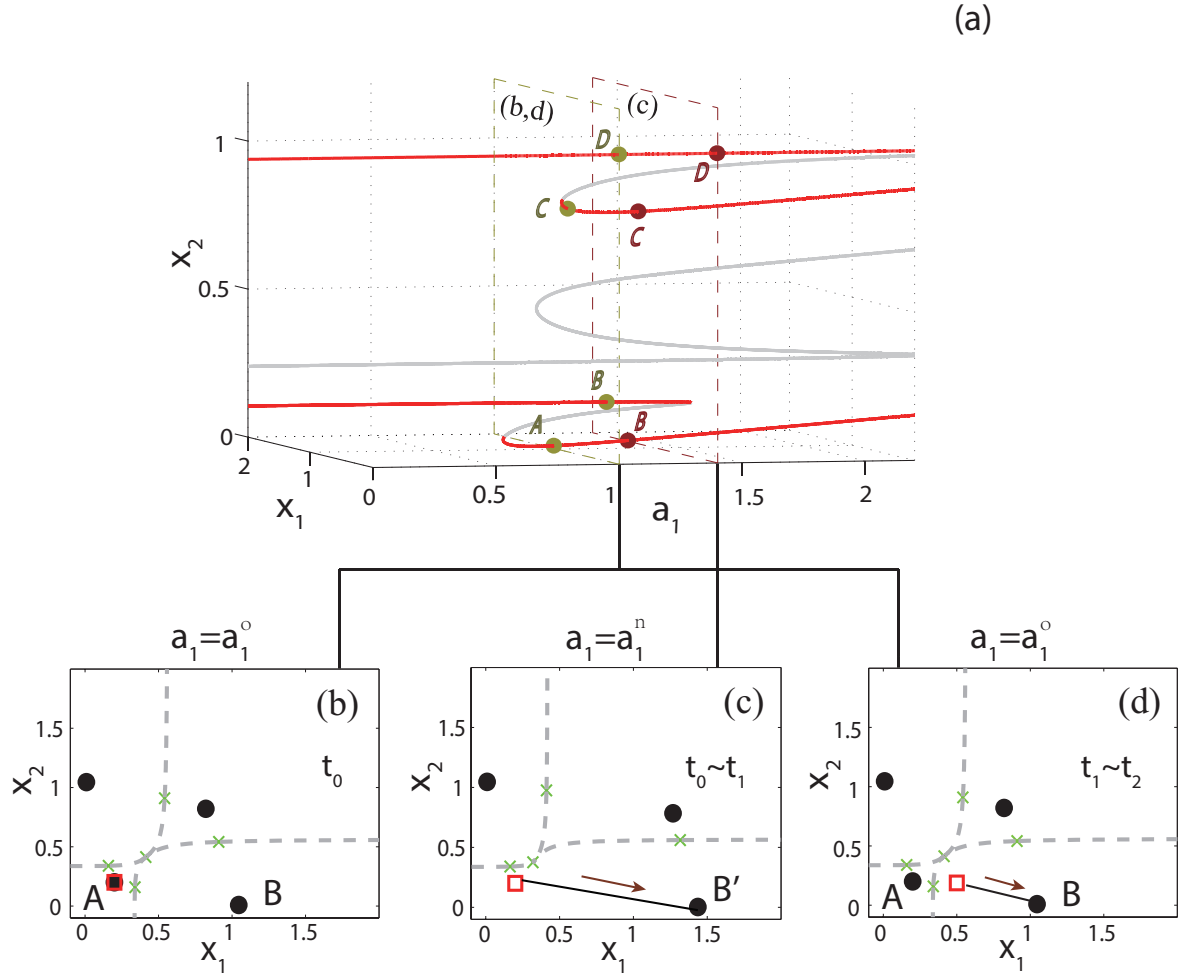,width=\linewidth}
\end{center}
\caption{\small {\bf Control of the two-node GRN.} 
(a) Bifurcation diagram with respect to the control parameter $a_1$, where 
the red and gray solid lines denote the stable and unstable steady states, 
respectively. In the two parallel cross-sections (with dashed line boundaries) 
for $a_1 = a_1^0$ and $a_1 = a_1^n$, the yellow and brown dots represent the 
corresponding stable attractors, respectively. In (b-d), gray dashed lines 
represent the basin boundaries; black solid circles and green crosses denote 
attractors and unstable steady states, respectively. (b) For the initial 
parameter setting, $a_1=a_1^0$, the system is at a low concentration 
state $\mathbf A$, and the target state is $\mathbf B$. (c) By changing 
$a_1$ from $a_1^0$ to $a_1^n$, attractor $\mathbf A$ is destabilized 
and its original basin is absorbed into that of the intermediate attractor 
${\mathbf B}^{\prime}$, so the system approaches ${\mathbf B}^{\prime}$. 
(d) When control perturbation upon $a_1$ is released, the landscape recovers 
to that in (b). Once the system has entered the basin of the target state 
$\mathbf B$ during the process in (c), it will evolve spontaneously towards 
$\mathbf B$. Parameters in simulation are $a_1^0=1.0$, $a_1^n=1.4$, $t_0=0$, 
$t_1=23$, and $t_2=40$.}
\label{fig:ctrl2D}
\end{figure*}

\begin{figure*}
\begin{center}
\epsfig{file=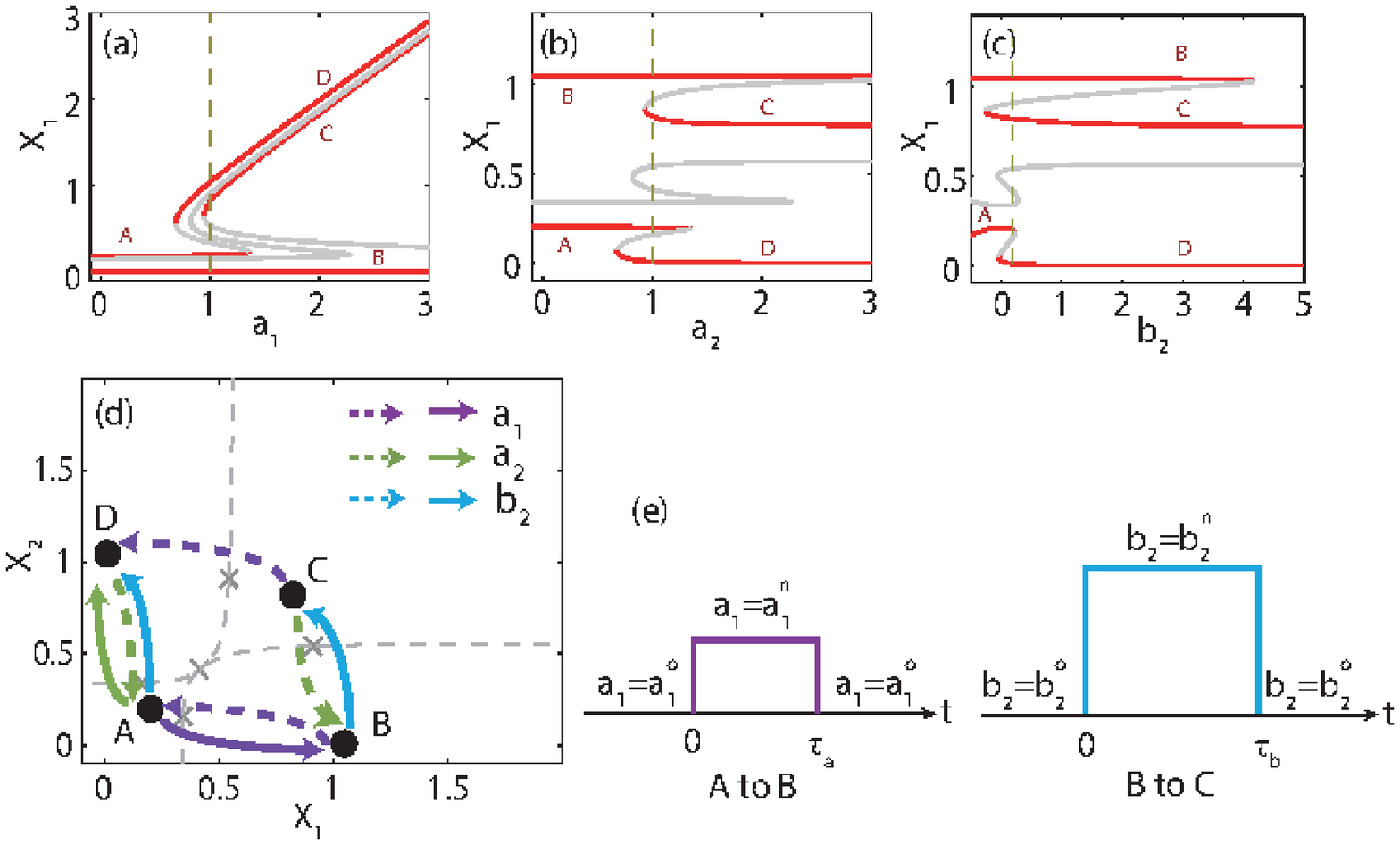,width=\linewidth}
\end{center}
\caption{\small {\bf Construction of attractor network for two-node GRN}.
(a-c) Bifurcation diagrams of the system with respect to the coupling 
parameters $a_1$, $a_2$ and $b_2$, respectively, where each bifurcation 
point can be exploited to design control. (d) The corresponding attractor 
network, in which each directed edge corresponds to an elementary control that 
is designed to steer the system from the original attractor to the directed 
one. The solid and dashed edges, respectively, denote the positive and 
negative changes in the corresponding control parameters. (e) Sequential 
control signals required to drive the system from attractor $\mathbf A$ 
to attractor $\mathbf C$ through the path 
$\mathbf A\rightarrow \mathbf B\rightarrow \mathbf C$. In simulation, the 
original parameter values are $a_1^0=1.0$ and $b_2^0=0.2$. We set 
$a_1^n=1.4$, followed by setting $b_2^n=4.2$, and the respective durations
of the parameter perturbation are $\tau_a=23$ and $\tau_b=32$.}
\label{fig:bifur2D}
\end{figure*}

In a typical experimental setting, four coupling parameters can be adjusted 
externally through the application of repressive or inductive drugs. To 
demonstrate attractor network and control implementation, we consider 
the parameter regime in which the system has four stable steady states 
(attractors) that correspond to four different cell states during cell 
development and differentiation. In particular, the dynamical network  
can be mathematically described as
\begin{equation} \label{eq:sys2D}
\begin{split}
\dot{x_1}=a_1 \cdot \frac{x_1^n}{s^n+x_1^n} +
b_1 \cdot \frac{s^n}{s^n+x_2^n}-k\cdot x_1, \\
\dot{x_2}=a_2 \cdot \frac{x_2^n}{s^n+x_2^n} +
b_2 \cdot \frac{s^n}{s^n+x_1^n}-k\cdot x_2, \\
\end{split}
\end{equation}
where the dynamical variables $(x_1, x_2)$ characterize the protein 
abundances of the genes products, $k$ denotes the degradation rate of 
each gene, and the tunable parameters $a_1$, $a_2$, $b_1$, and $b_2$ represent 
the strengths of auto or mutual regulations. In a GRN, the dynamical behaviors
of inhibition and activation are captured by the Hill function: 
$f(x)=x^n/(x^n + s^n)$ for activation and $f(x)=s^n/(x^n +s^n)$ for inhibition,
where the parameter $s$ characterizes half activation (or inhibition) 
concentration (for $x = s$, the output is $0.5$), and $n$ quantifies the 
correlation between the input and output concentrations, where a larger 
value of $n$ corresponds to a stronger inhibition or activation effect. 
For any specific GRN, the values of both $s$ and $n$ can be determined 
experimentally. For simplicity, we assume the system to be 
symmetric in that the inhibition and activation share the same Hill function
(i.e., with the same parameters $s$ and $n$). To have four attractors, we set 
the auto activation strengths, $a_1$ and $a_2$, to $1.0$, and mutual 
inhibition strengths, $b_1$ and $b_2$, to $0.2$. The value of the 
degradation rate $k$ is set to $1.1$, taking into account the effects of 
protein degradation and cell volume expansion.

\begin{figure*}
\begin{center}
\epsfig{file=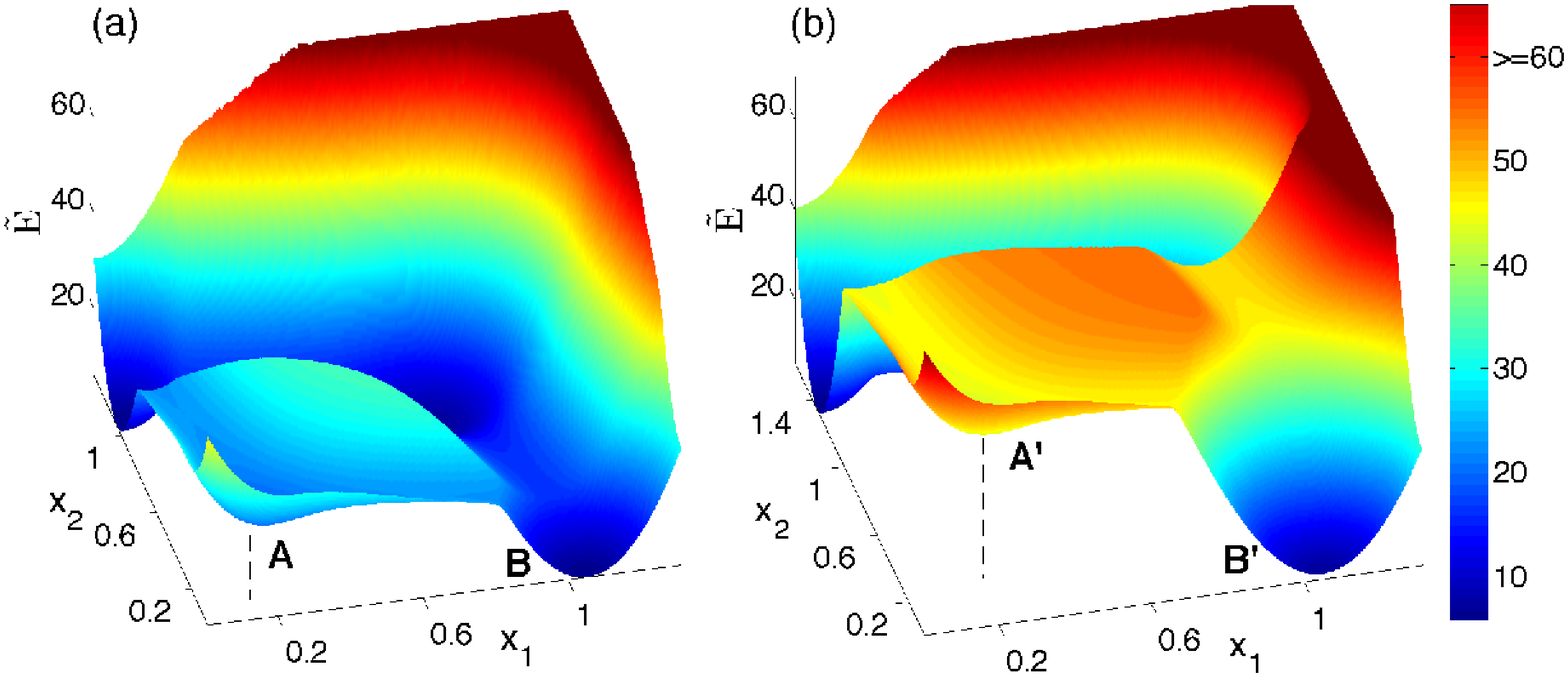,width=\linewidth} 
\caption{\small {\bf Illustration of pseudo potential landscape}.
``Pseudo'' potential $\tilde{E}$ of the two-node GRN system (\textbf{a}) 
for $a_1=1.0$ ($\Delta_\mathrm{d} \approx 0.3549$), $\sigma=0.05$ 
and (\textbf{b}) for $a_1=1.3$ ($\Delta_\mathrm{d} \approx 0.0549$), 
$\sigma=0.05$. Regions of warm and cold colors indicate the states with 
large and small pseudo energies, respectively.} 
\label{fig:landscape}
\end{center}
\end{figure*}

Figure~\ref{fig:ctrl2D} shows a particular process of controlling the 
system from an initial state, denoted as ${\mathbf A}$, in which both 
$x_1$ and $x_2$ have low abundance, to a final state ${\mathbf B}$ where 
$x_1$ and $x_2$ have high and low abundance, respectively. From the 
bifurcation diagram [Fig.~\ref{fig:ctrl2D}(a)] with respect to the control 
parameter $a_1$, we see that, as $a_1$ is increased from $1.0$ to $1.4$,
in the lower branch, the initial attractor $\mathbf A$ is destabilized 
through a saddle-node bifurcation. The bifurcation-based control process 
is shown in Figs.~\ref{fig:ctrl2D}(b-d), where panel (b) exhibits 
the phase space of the system prior to control ($a_1=1.0$). When control is 
activated so that $a_1$ is set to $a_1=1.4$, the original basin of 
attraction of attractor ${\mathbf A}$ merges into the basin of an 
intermediate attractor ${\mathbf B}^{\prime}$, and the system originally
in ${\mathbf A}$ starts to migrate towards the 
intermediate attractor ${\mathbf B}^{\prime}$, as indicated by the arrowed 
trajectory in panel (c). Control perturbation upon $a_1$ can be withdrawn 
once the state of the system enters the region belonging to the original 
basin of the target attractor $\mathbf B$, after which the system  
spontaneously evolves into $\mathbf B$ for $a_1=1.0$, as shown in 
Fig.~\ref{fig:ctrl2D}(d). 

To obtain a global picture of all possible control outcomes, we construct 
the attractor network for the two-node GRN system, assuming that three 
control parameters: $a_1$, $a_2$ and $b_2$, are available for control. The 
corresponding bifurcation diagrams are shown in Figs.~\ref{fig:bifur2D}(a-c), 
from which all saddle-node bifurcations can be identified for control
design. When all the attractors are connected with directed and weighted 
edges through the control processes, i.e., when none of the attractor
is isolated, we obtain an attractor network, as shown in 
Fig.~\ref{fig:bifur2D}(d). Specifically, the edge weight can be assigned
by taking into account the key characteristics of control such as the critical 
parameter strength $\mu_c$ and the power-law scaling behavior of the 
required minimum control time $\tau_m$ (see Supplementary Table~2).
From the attractor network, we can find all possible control paths for any 
given pair of original and desired states. 

From Fig.~\ref{fig:bifur2D}(d), we see that the two-node GRN system is 
\emph{fully controllable} since any of the coexisting attractors is 
reachable by applying proper sequential controls upon the 
available parameters. The concept of attractor network is appealing because
it provides a clear control scenario to drive the system from any initial
attractor to any desired attractor. In fact, the attractor network provides
a blueprint that can be used to design a proper combination of parameter 
changes to induce the so-called synergistic or antagonistic 
effects~\cite{Yin:2014}. For example, $\mathbf A$ is not directly connected
with $\mathbf C$, neither is $\mathbf B$ directly connected to $\mathbf D$.
However, the system can be steered from $\mathbf A$ to $\mathbf B$ through 
perturbation on $a_1$, and then from $\mathbf B$ to $\mathbf C$ through 
parameter perturbation on $b_2$, as shown in Fig.~\ref{fig:bifur2D}(e).
Another example to demonstrate the need of multiple parameter perturbation
is to control the system from $\mathbf B$ to $\mathbf D$. A viable control 
path is $\mathbf B\rightarrow \mathbf C\rightarrow \mathbf D$, which can be 
realized through perturbation on parameters $(b_2, a_1)$. We also see that 
the two $\mathbf B\rightarrow \mathbf A\rightarrow \mathbf D$ paths can be 
realized through parameter changes in $(a_1, a_2)$ and $(a_1, b_2)$, 
respectively. 

When multiple control paths exist from an initial attractor to a final one,
a practical issue is to identify an optimal path that is cost effective
and robust. The concept of \emph{weighted-shortest path} can be used to 
address this issue. Particularly, the weights of edges can be determined
from experimental considerations such as the cost, limitation in drug dose, 
the control duration time, etc. 

\paragraph*{Potential landscape and beneficial role of noise in nonlinear
control.} The role of noise in facilitating control of a nonlinear 
dynamical network can be understood using the concept of 
{\em potential landscape}~\cite{WZXW:2011,BZA:2011,KST:2013}
or Waddington landscape~\cite{Waddington:book} in systems 
biology, which essentially determines the biological paths for cell 
development and differentiation~\cite{Huang:2009,MML:2009,CS:2012} - the 
{\em landscape metaphor}. The potential landscape has been used to 
manipulate time scales to control stochastic and induced switching 
in biophysical networks~\cite{CS:2012}. Intuitively, 
the power of the concept of the landscape can be understood by resorting 
to the elementary physical picture of a ball moving in a valley under 
gravity. The valley thus corresponds to one stable attractor. To the 
right of the valley there is a hill, or a potential barrier in the 
language of classical mechanics. The downhill side to 
the right of the barrier corresponds to a different attractor. Suppose
the confinement of ball's motion within the valley is undesired and one
wishes to push the ball over the barrier to the right attractor (desired).
If the valley is deep (or the height of the barrier is large), 
there will be little probability for the ball to move across the top 
of the barrier towards the desired attractor. In this case, a small amount 
of noise is unable to enhance the crossover probability. However, if the 
barrier height is small, a small amount of noise can 
push the ball over to desired attractor on the right side of the barrier.
Thus, the beneficial role of noise is more pronounced for small height of
the potential barrier, a behavior that we observe when controlling the
T-cell network (Fig.~\ref{fig:uc_noiseTcell}). In mechanics, the system 
can be formulated using a potential function so that, mathematically, the 
motion of the ball can be described by the Langevin equation, which has been 
a paradigmatic model to understand nonlinear phenomena such as stochastic 
resonance~\cite{BSV:1981,BPSV:1983,MW:1989,MPO:1994,GNCM:1997,GHJM:1998}. 
In the past few years, a quantitative
approach has been developed to mapping out the potential landscape for gene
circuits or gene regulatory networks~\cite{WXW:2008,WXWH:2010,WZXW:2011,
ZXZWW:2012}. In nonlinear dynamical systems, a similar concept exists -
{\em quasipotential}~\cite{GT:1984,GHT:1991,TL:2010}, which plays 
an important role in understanding phenomena such as noise-induced chaos.


For an attractor network, in the presence of noise each node corresponds to 
a potential valley of certain depth characterizing the stability of the 
attractor. For a fixed depth, noise of larger amplitude $\sigma$ leads  
a larger escaping probability or shorter escaping time. When the amplitude
of the control signal is not sufficient to drive the system across the 
local potential barrier, noise can facilitate control by pushing the 
system out of the undesired valley (attractor).   

The potential landscape for a GRN under Gaussian noise can be constructed
from the dynamical equations of the system using the concept of ``pseudo''
energy~\cite{KST:2013} (see {\bf Methods}). For the two-node GRN system 
[Eq.~(\ref{eq:sys2D})] subject to stochastic disturbance $N(0,\sigma^2)$,
we can compute the potential landscape for any combination of a system 
parameter (say $a_1$) and the noise amplitude $\sigma$. 
Figure~\ref{fig:landscape} shows two examples of the landscapes for 
$a_1=1.0$ and $a_1=1.3$, where the noise amplitude is $\sigma=0.05$. 
We see that, for $a_1 = 1.0$, there are four valleys (attractors).
For $a_1 = 1.3$, the pseudo energy for $\mathbf A$ (the original valley at 
the lower-left corner) becomes higher, and the path for the transition 
from $\mathbf A$ to $\mathbf B$ becomes more pronounced. Further increasing 
$a_1$ towards the critical value (about $1.35$) raises the energy of 
$\mathbf A$ to the level of the potential barrier, effectively eliminating 
the corresponding valley and the attractor itself. 

\begin{figure*}[h]
\begin{center}
\epsfig{file=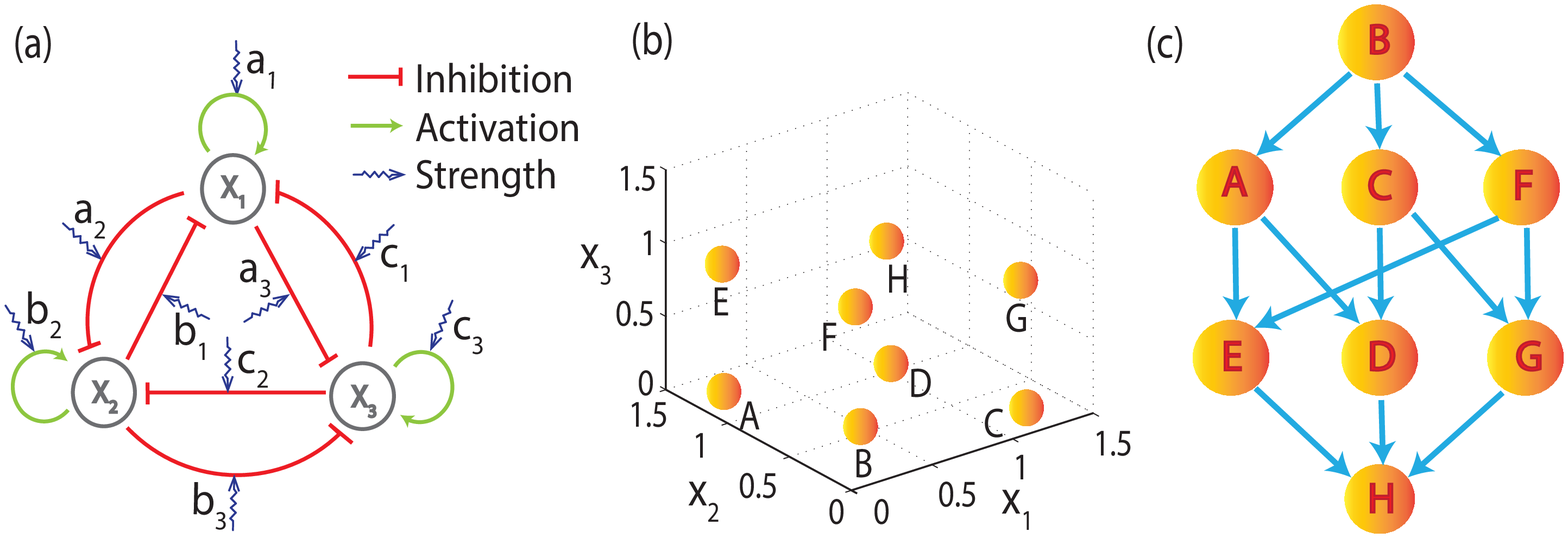,width=1\linewidth}
\end{center}
\caption{ {\bf Attractor network and control of a three-node GRN system.}
(a) Schematic illustration of a three-node GRN system. The arrowhead and
bar-head edges represent activation and inhibitory regulations, respectively.
The sawtooth lines specify that corresponding edge strength is experimentally
adjustable. (b) Coexisting attractors ($\mathbf A$ to $\mathbf H$) in the
phase space. (c) The underlying attractor network, where each node
represents an attractor and each weighted directed link indicates that
its strength can be experimentally tuned to steer the system from the
starting attractor to the pointed attractor.}
\label{fig:3D}
\end{figure*}

\paragraph*{Attractor network for a three-node GRN.}
We also study a three-node GRN system, as shown in Fig.~\ref{fig:3D}(a).
Similar to the two-node GRN system, there exist both auto and mutual
regulations among the nodes. All the interactions are assumed to be
characterized by the same parameters of $s$ and $n$ in the Hill function. The 
nonlinear dynamical equations of the system are~\cite{MTELT:2009,Shuetal:2013}
\begin{equation} \label{eq:sys3D}
\begin{split}
\dot{x_1}=a_1 \cdot \frac{x_1^n}{s^n+x_1^n}+b_1 \cdot \frac{s^n}{s^n+x_2^n}+c_1 \cdot \frac{s^n}{s^n+x_3^n}-k\cdot x_1, \\
\dot{x_2}=a_2 \cdot \frac{s^n}{s^n+x_1^n}+b_2 \cdot \frac{x_2^n}{s^n+x_2^n}+c_2 \cdot \frac{s^n}{s^n+x_3^n}-k\cdot x_2, \\
\dot{x_3}=a_3 \cdot \frac{s^n}{s^n+x_1^n}+b_3 \cdot \frac{s^n}{s^n+x_2^n}+c_3 \cdot \frac{x_3^n}{s^n+x_3^n}-k\cdot x_3, \\
\end{split}
\end{equation}
where the state variables ($x_1$, $x_2$ and $x_3$) represent the
abundances of the three genes products, the auto-activation parameters
$a_1$, $b_2$, $c_3$ and the mutual-inhibition parameters $a_2$, $a_3$,
$b_1$, $b_3$, $c_1$, $c_2$ are all experimentally accessible.
To be concrete, initially all the auto activation and mutual inhibition
parameters are set to be $1.0$ and $0.1$, respectively, and $k$ is the
degradation rate that can be conveniently set to unity. The parameters
in the Hill function are $n = 4$ and $s = 0.5$. There are altogether
eight attractors in this system, as shown in Fig.~\ref{fig:3D}(b), which
are distributed symmetrically in the three-dimensional state space. For
example, attractor $\mathbf H$ has relatively high values for all
three dynamical variables, and attractor $\mathbf B$ exhibits the
opposite case with low abundances. For attractors $\mathbf A$,
$\mathbf C$ and $\mathbf F$, one of the three state variables is high
and the other two are low. For attractors $\mathbf D$, $\mathbf E$
and $\mathbf G$, one of the three state variables is low but the other
two are high.

From numerical simulations, we find that the features of control are
essentially the same as those for the two-node GRN system, in terms of
characteristics such as the existence of critical control strength and
the power-law behavior of the minimum control time (see
Table~S3 in SI). We construct the attractor network
in Fig.~\ref{fig:3D}(c) through combinations of all eight attractors (as
nodes) and directed elementary controls (as weighted directed edges).
Information in Table~S3 can also be used to
estimate the respective weights of the edges. From the attractor network,
for any given pair of initial and final states, we can identify all the
viable control paths. Furthermore, the \emph{weighted-shortest path}
can be calculated once the edge weights are determined.

We note that, typically, the attractor network based on elementary control
is not an all-to-all directed network so that certain control paths are
absent, e.g., from attractor $\mathbf H$ to $\mathbf B$. The biological
meaning is that, while a stem state can be differentiated into various
types of cells through bifurcation, the opposite paths back to the stem
state are much more difficult to find. \\
\\ \noindent
{\large\bf Discussions} \\
\\ \noindent
The field of controlling chaos in nonlinear dynamical systems has been 
active for more than two decades since the seminal work of Ott, Grebogi,
and Yorke~\cite{OGY:1990}. The basic idea was that chaos, while
signifying random or irregular behavior, possesses an intrinsically
sensitive dependence on initial conditions, which can be exploited for 
controlling the system using only small perturbations. This feature,
in combination with the fact that a chaotic system possesses an infinite set
of unstable periodic orbits, each leading to different system performance, 
implies that a chaotic system can be stabilized about some desired 
state with optimal performance using small control perturbations. 
Controlling chaos has since been studied extensively and examples of 
successful experimental implementation abound in physical, chemical, 
biological, and engineering systems~\cite{BGLMM:2000}. The vast 
literature on controlling chaos, however, has been mostly limited to 
low-dimensional systems, systems that possess one or a very few 
unstable directions (i.e., one or a very few positive Lyapunov exponents). 
Complex networks with nonlinear dynamics are generally high dimensional, 
rendering inapplicable existing methodologies of chaos control. While
mathematical frameworks of controllability 
of complex networks~\cite{LSB:2011,YZDWL:2013}
were developed and extensively studied recently, the deficiency of such 
rigorous mathematical frameworks is that the nodal 
dynamical processes must be assumed to be linear. 
While controllability for nonlinear control can be formulated based on
Lie brackets~\cite{SL:book}, it may be difficult to implement the abstract
framework for complex networks. 

Controlling nonlinear dynamics on complex networks remains to be an 
outstanding and challenging problem, especially in terms of the two
key issues: controllability and actual control. To assess the 
controllability of nonlinear dynamical networks, drastically different
approaches than the linear controllability framework are needed.
While there were previous works on specific control methods such as pinning
control~\cite{WC:2002,LWC:2004,SBGC:2007} and brute-force control that 
rely on altering the state variables of the underlying system (which
in realistic situations can be difficult to implement), we continue to 
lack a general framework for actual control of complex networks with 
nonlinear dynamics through realistic, physical means. The main difficulty 
lies in the extremely diverse nonlinear dynamical behaviors 
that a network can generate, making it practically impossible to develop
a general mathematical framework for control. In particular, the 
traditional control theoretical tools for linear dynamical systems aim to
control the detailed states of all of the variables, which is in
fact an overkill for most systems. For nonlinear dynamical networks, 
a physically meaningful approach may not require detailed control of all state 
variables. With this relaxation of the control requirement, it may be
possible to develop a framework of controllability and devise actual
control strategies for nonlinear dynamical networks based on 
physical/experimental considerations.

A common feature of nonlinear
dynamical systems is the emergence of a large number of distinct,
coexisting attractors~\cite{Ott:book,LT:book}. Often the performance and
functions of the system are determined by the particular attractor that the 
system has settled into, associated with which the detailed states of the 
dynamical variables are not relevant. The key is thus to develop control 
principles whereby we nudge a complex, nonlinear system from attractor to
attractor through small perturbations to a set of physically or 
experimentally feasible parameters. The main message of this paper
is that a controllability framework can be developed for nonlinear 
dynamical networks based on the control of attractors.

Generally, the reason for control is that the current system is likely to 
evolve into an undesired state (attractor) or the system is already
in such a state, and one wishes to apply perturbations to bring the
system out of the undesired state and steer it into a desired state.
The first step is then to identify a final state or attractor of the system
that leads to the desired performance. The next step is to choose a
set of experimentally adjustable parameters and determine whether small
perturbations to these parameter can bring the system to the desired
attractor. That is, under physically realizable perturbations there
should be a control path between the undesired and the desired attractors.
The path can be directly from the former to the latter, or there can
be intermediate attractors on the path. For example, due to the physical
constraint on the control parameters and the ranges in which they can be 
meaningfully varied, one can drive the system into some intermediate attractors
by perturbing one set of parameters, and then from these attractors
to the final attractor by using a different set of parameter control.
For a complex, nonlinear dynamical network, the number of coexisting
attractors can be large. Given a set of system performance indicators,
one can classify all the available attractors into three categories:
the undesired, desired, and the intermediate attractors. We say
a nonlinear network is controllable if there is a control path from
any undesired attractor to the desired attractor under finite
parameter perturbations. Regarding each attractor as a node and the
control paths as directed links or edges, we generate an
attractor network whose properties determine the controllability
of the original networked dynamical system. For example, the average
path length from an undesired to a desired attractor and
the ``control energy'' (or the amount of necessary parameter
perturbations) can serve as quantitative measures to characterize
the controllability of the original network. We demonstrate our 
idea of control and construction of attractor networks using realistic
networks from systems and synthetic biology. We also find that noise
can facilitate control of nonlinear dynamical networks, and we provide
a physical theory to understand this counterintuitive phenomenon.

While we emphasize the need to focus on physically meaningful and 
experimentally accessible parameter perturbations, there can still be 
a large number of attractor networks depending on the choice of the 
parameters, making it difficult to formulate a rigorous mathematical 
framework. We believe that these issues can and will be satisfactorily 
addressed in the near future, ultimately realizing the grand goal of 
controlling nonlinear dynamical networks. \\
\\ \noindent
{\large\bf Methods} \\ 

\paragraph*{Pseudo potential landscape.}
For a dissipative, nonlinear dynamical system subject to noise, we can 
construct a pseudo potential landscape based on the state probability 
distribution. Assume that, asymptotically, the system approaches a 
stationary distribution. For a canonical dynamical system, the potential
can be defined as $E({\bf x}) = - \log{P({\bf x})}$, where $P({\bf x})$
is the probability density function. For a conservative dynamical system, 
the direction of system evolution is nothing but the direction of the 
gradient of the potential function. However, this does not hold for 
dissipative dynamical systems. The potential function thus does not 
have the same physical meaning as that for a conservative system, 
henceforth the term pseudo potential. This approach can be adopted
to GRNs.  

To obtain the stationary distribution, we use the modified
weighted-ensemble algorithm proposed by Kromer et al.~\cite{KST:2013},
which offers faster convergence than, for example, the traditional
random walk method. To be illustrative, we take the two-node GRN system
[Eq.~(\ref{eq:sys2D})] as an example to demonstrate how the pseudo 
potential landscape can be numerically obtained. The state space of the 
two-dimensional dynamical system is partitioned into an \textbf{$M\times M$} 
lattice with reflective boundaries conditions. Initially the probability 
$P_{m,n}(t)$ of all gird points are set to be uniform. The simulation time 
is divided into $T$ steps, where each step has the duration $\tau$. At the 
beginning of each step $t$, there are $N$ walkers randomly distributed at 
the grid point $(m,n)$, which carry equal weight $P_{m,n}(t)/N$ and perform 
random walk under the system dynamics and noise. The locations of these walkers 
in the grid are recorded at the end of each time step, and the probability 
at next time step, $P_{m,n}(t+1)$, is the summation of the probabilities 
carried by all the walkers at time $t$. At time $(t+1)$, $N$ new walkers
carrying the updated probability at each grid point perform random walk 
again on the grid. This procedure repeats until the probability distribution 
becomes stationary, say $P_{m,n}$, which gives the pseudo potential landscape 
as $\tilde{E}(m,n)=-\log{P_{m,n}}$. Numerically, the time evolution of all 
walkers can be simulated using the second-order Heun method for integrating
stochastic differential equations. For Fig.~\ref{fig:landscape}, the 
state space is divided into a $500\times 500$ grid. At each grid point
there are $N=20$ walkers, each evolving $T=2000$ time steps with 
$\tau=10^{-4}$. 


\begin{thebibliography}{10}
\expandafter\ifx\csname url\endcsname\relax
  \def\url#1{\texttt{#1}}\fi
\expandafter\ifx\csname urlprefix\endcsname\relax\def\urlprefix{URL }\fi
\providecommand{\bibinfo}[2]{#2}
\providecommand{\eprint}[2][]{\url{#2}}

\bibitem{LH:2007}
\bibinfo{author}{Lombardi, A.} \& \bibinfo{author}{H\"{o}rnquist, M.}
\newblock \bibinfo{title}{Controllability analysis of networks}.
\newblock \emph{\bibinfo{journal}{Phys. Rev. E}} \textbf{\bibinfo{volume}{75}},
  \bibinfo{pages}{056110} (\bibinfo{year}{2007}).

\bibitem{LCWX:2008}
\bibinfo{author}{Liu, B.}, \bibinfo{author}{Chu, T.}, \bibinfo{author}{Wang,
  L.} \& \bibinfo{author}{Xie, G.}
\newblock \bibinfo{title}{Controllability of a leader-follower dynamic network
  with switching topology}.
\newblock \emph{\bibinfo{journal}{IEEE Trans. Automat. Contr.}}
  \textbf{\bibinfo{volume}{53}}, \bibinfo{pages}{1009--1013}
  (\bibinfo{year}{2008}).

\bibitem{RJME:2009}
\bibinfo{author}{Rahmani, A.}, \bibinfo{author}{Ji, M.},
  \bibinfo{author}{Mesbahi, M.} \& \bibinfo{author}{Egerstedt, M.}
\newblock \bibinfo{title}{Controllability of multi-agent systems from a
  graph-theoretic perspective}.
\newblock \emph{\bibinfo{journal}{SIAM J. Contr. Optim.}}
  \textbf{\bibinfo{volume}{48}}, \bibinfo{pages}{162--186}
  (\bibinfo{year}{2009}).

\bibitem{LSB:2011}
\bibinfo{author}{Liu, Y.-Y.}, \bibinfo{author}{Slotine, J.-J.} \&
  \bibinfo{author}{Barab\'asi, A.-L.}
\newblock \bibinfo{title}{Controllability of complex networks}.
\newblock \emph{\bibinfo{journal}{Nature (London)}}
  \textbf{\bibinfo{volume}{473}}, \bibinfo{pages}{167--173}
  (\bibinfo{year}{2011}).

\bibitem{WNLG:2011}
\bibinfo{author}{Wang, W.-X.}, \bibinfo{author}{Ni, X.}, \bibinfo{author}{Lai,
  Y.-C.} \& \bibinfo{author}{Grebogi, C.}
\newblock \bibinfo{title}{Optimizing controllability of complex networks by
  small structural perturbations}.
\newblock \emph{\bibinfo{journal}{Phys. Rev. E}} \textbf{\bibinfo{volume}{85}},
  \bibinfo{pages}{026115} (\bibinfo{year}{2011}).

\bibitem{YZDWL:2013}
\bibinfo{author}{Yuan, Z.-Z.}, \bibinfo{author}{Zhao, C.}, \bibinfo{author}{Di,
  Z.-R.}, \bibinfo{author}{Wang, W.-X.} \& \bibinfo{author}{Lai, Y.-C.}
\newblock \bibinfo{title}{Exact controllability of complex networks}.
\newblock \emph{\bibinfo{journal}{Nature Commun.}}
  \textbf{\bibinfo{volume}{4}}, \bibinfo{pages}{2447} (\bibinfo{year}{2013}).

\bibitem{CWWL:2015}
\bibinfo{author}{Chen, Y.-Z.}, \bibinfo{author}{Wang, L.-Z.},
  \bibinfo{author}{Wang, W.-X.} \& \bibinfo{author}{Lai, Y.-C.}
\newblock \bibinfo{title}{The paradox of controlling complex networks: control
  inputs versus energy requirement}.
\newblock \emph{\bibinfo{journal}{arXiv:1509.03196v1}}  (\bibinfo{year}{2015}).

\bibitem{NA:2012}
\bibinfo{author}{Nacher, J.~C.} \& \bibinfo{author}{Akutsu, T.}
\newblock \bibinfo{title}{Dominating scale-free networks with variable scaling
  exponent: heterogeneous networks are not difficult to control}.
\newblock \emph{\bibinfo{journal}{New J. Phys.}} \textbf{\bibinfo{volume}{14}},
  \bibinfo{pages}{073005} (\bibinfo{year}{2012}).

\bibitem{YRLLL:2012}
\bibinfo{author}{Yan, G.}, \bibinfo{author}{Ren, J.}, \bibinfo{author}{Lai,
  Y.-C.}, \bibinfo{author}{Lai, C.-H.} \& \bibinfo{author}{Li, B.}
\newblock \bibinfo{title}{Controlling complex networks: How much energy is
  needed?}
\newblock \emph{\bibinfo{journal}{Phys. Rev. Lett.}}
  \textbf{\bibinfo{volume}{108}}, \bibinfo{pages}{218703}
  (\bibinfo{year}{2012}).

\bibitem{NV:2012}
\bibinfo{author}{Nepusz, T.} \& \bibinfo{author}{Vicsek, T.}
\newblock \bibinfo{title}{Controlling edge dynamics in complex networks}.
\newblock \emph{\bibinfo{journal}{Nat. Phys.}} \textbf{\bibinfo{volume}{8}},
  \bibinfo{pages}{568--573} (\bibinfo{year}{2012}).

\bibitem{LSBA:2013}
\bibinfo{author}{Liu, Y.-Y.}, \bibinfo{author}{Slotine, J.-J.} \&
  \bibinfo{author}{Barab{\'a}si, A.-L.}
\newblock \bibinfo{title}{Observability of complex systems}.
\newblock \emph{\bibinfo{journal}{Proc. Natl. Acad. Sci. (USA)}}
  \textbf{\bibinfo{volume}{110}}, \bibinfo{pages}{2460--2465}
  (\bibinfo{year}{2013}).

\bibitem{MDB:2014}
\bibinfo{author}{Menichetti, G.}, \bibinfo{author}{Dall¡¯Asta, L.} \&
  \bibinfo{author}{Bianconi, G.}
\newblock \bibinfo{title}{Network controllability is determined by the density
  of low in-degree and out-degree nodes}.
\newblock \emph{\bibinfo{journal}{Phys. Rev. Lett.}}
  \textbf{\bibinfo{volume}{113}}, \bibinfo{pages}{078701}
  (\bibinfo{year}{2014}).

\bibitem{RR:2014}
\bibinfo{author}{Ruths, J.} \& \bibinfo{author}{Ruths, D.}
\newblock \bibinfo{title}{Control profiles of complex networks}.
\newblock \emph{\bibinfo{journal}{Science}} \textbf{\bibinfo{volume}{343}},
  \bibinfo{pages}{1373--1376} (\bibinfo{year}{2014}).

\bibitem{Wuchty:2014}
\bibinfo{author}{Wuchty, S.}
\newblock \bibinfo{title}{Controllability in protein interaction networks}.
\newblock \emph{\bibinfo{journal}{Proc. Natl. Acad. Sci. (USA)}}
  \textbf{\bibinfo{volume}{111}}, \bibinfo{pages}{7156--7160}
  (\bibinfo{year}{2014}).

\bibitem{WBSS:2015}
\bibinfo{author}{Whalen, A.~J.}, \bibinfo{author}{Brennan, S.~N.},
  \bibinfo{author}{Sauer, T.~D.} \& \bibinfo{author}{Schiff, S.~J.}
\newblock \bibinfo{title}{Observability and controllability of nonlinear
  networks: The role of symmetry}.
\newblock \emph{\bibinfo{journal}{Phys. Rev. X}} \textbf{\bibinfo{volume}{5}},
  \bibinfo{pages}{011005} (\bibinfo{year}{2015}).

\bibitem{YTBSLB:2015}
\bibinfo{author}{Yan, G.} \emph{et~al.}
\newblock \bibinfo{title}{Spectrum of controlling and observing complex
  networks}.
\newblock \emph{\bibinfo{journal}{Nat. Phys.}} \textbf{\bibinfo{volume}{11}},
  \bibinfo{pages}{779--786} (\bibinfo{year}{2015}).

\bibitem{Kalman:1963}
\bibinfo{author}{Kalman, R.~E.}
\newblock \bibinfo{title}{Mathematical description of linear dynamical
  systems}.
\newblock \emph{\bibinfo{journal}{J. Soc. Indus. Appl. Math. Ser. A}}
  \textbf{\bibinfo{volume}{1}}, \bibinfo{pages}{152--192}
  (\bibinfo{year}{1963}).

\bibitem{Lin:1974}
\bibinfo{author}{Lin, C.-T.}
\newblock \bibinfo{title}{Structural controllability}.
\newblock \emph{\bibinfo{journal}{IEEE Trans. Automat. Contr.}}
  \textbf{\bibinfo{volume}{19}}, \bibinfo{pages}{201--208}
  (\bibinfo{year}{1974}).

\bibitem{Luenberger:book}
\bibinfo{author}{Luenberger, D.~G.}
\newblock \emph{\bibinfo{title}{Introduction to Dynamical Systems: Theory,
  Models, and Applications}} (\bibinfo{publisher}{John Wiley \& Sons, Inc},
  \bibinfo{address}{New Jersey}, \bibinfo{year}{1999}),
  \bibinfo{edition}{first} edn.

\bibitem{WC:2002}
\bibinfo{author}{Wang, X.~F.} \& \bibinfo{author}{Chen, G.}
\newblock \bibinfo{title}{Pinning control of scale-free dynamical networks}.
\newblock \emph{\bibinfo{journal}{Physica A}} \textbf{\bibinfo{volume}{310}},
  \bibinfo{pages}{521--531} (\bibinfo{year}{2002}).

\bibitem{LWC:2004}
\bibinfo{author}{Li, X.}, \bibinfo{author}{Wang, X.~F.} \&
  \bibinfo{author}{Chen, G.}
\newblock \bibinfo{title}{Pinning a complex dynamical network to its
  equilibrium}.
\newblock \emph{\bibinfo{journal}{IEEE Trans. Circ. Syst. I}}
  \textbf{\bibinfo{volume}{51}}, \bibinfo{pages}{2074--2087}
  (\bibinfo{year}{2004}).

\bibitem{SBGC:2007}
\bibinfo{author}{Sorrentino, F.}, \bibinfo{author}{di~Bernardo, M.},
  \bibinfo{author}{Garofalo, F.} \& \bibinfo{author}{Chen, G.}
\newblock \bibinfo{title}{Controllability of complex networks via pinning}.
\newblock \emph{\bibinfo{journal}{Phys. Rev. E}} \textbf{\bibinfo{volume}{75}},
  \bibinfo{pages}{046103} (\bibinfo{year}{2007}).

\bibitem{CZL:2014}
\bibinfo{author}{Chen, Y.-Z.}, \bibinfo{author}{Huang, Z.-G.} \&
  \bibinfo{author}{Lai, Y.-C.}
\newblock \bibinfo{title}{Controlling extreme events on complex networks}.
\newblock \emph{\bibinfo{journal}{Sci. Rep.}} \textbf{\bibinfo{volume}{4}},
  \bibinfo{pages}{6121} (\bibinfo{year}{2014}).

\bibitem{GMOY:1983}
\bibinfo{author}{Grebogi, C.}, \bibinfo{author}{McDonald, S.~W.},
  \bibinfo{author}{Ott, E.} \& \bibinfo{author}{Yorke, J.~A.}
\newblock \bibinfo{title}{Final state sensitivity: an obstruction to
  predictability}.
\newblock \emph{\bibinfo{journal}{Phys. Lett. A}}
  \textbf{\bibinfo{volume}{99}}, \bibinfo{pages}{415--418}
  (\bibinfo{year}{1983}).

\bibitem{MGOY:1985}
\bibinfo{author}{McDonald, S.~W.}, \bibinfo{author}{Grebogi, C.},
  \bibinfo{author}{Ott, E.} \& \bibinfo{author}{Yorke, J.~A.}
\newblock \bibinfo{title}{Fractal basin boundaries}.
\newblock \emph{\bibinfo{journal}{Physica D}} \textbf{\bibinfo{volume}{17}},
  \bibinfo{pages}{125--153} (\bibinfo{year}{1985}).

\bibitem{FG:1997}
\bibinfo{author}{Feudel, U.} \& \bibinfo{author}{Grebogi, C.}
\newblock \bibinfo{title}{Multistability and the control of complexity}.
\newblock \emph{\bibinfo{journal}{Chaos}} \textbf{\bibinfo{volume}{7}},
  \bibinfo{pages}{597--604} (\bibinfo{year}{1997}).

\bibitem{FG:2003}
\bibinfo{author}{Feudel, U.} \& \bibinfo{author}{Grebogi, C.}
\newblock \bibinfo{title}{Why are chaotic attractors rare in multistable
  systems?}
\newblock \emph{\bibinfo{journal}{Phys. Rev. Lett.}}
  \textbf{\bibinfo{volume}{91}}, \bibinfo{pages}{134102}
  (\bibinfo{year}{2003}).

\bibitem{LT:book}
\bibinfo{author}{Lai, Y.-C.} \& \bibinfo{author}{T\'el, T.}
\newblock \emph{\bibinfo{title}{Transient Chaos - Complex Dynamics on
  Finite-Time Scales}} (\bibinfo{publisher}{Springer}, \bibinfo{address}{New
  York}, \bibinfo{year}{2011}), \bibinfo{edition}{first} edn.

\bibitem{NYLDG:2013}
\bibinfo{author}{Ni, X.}, \bibinfo{author}{Ying, L.}, \bibinfo{author}{Lai,
  Y.-C.}, \bibinfo{author}{Do, Y.-H.} \& \bibinfo{author}{Grebogi, C.}
\newblock \bibinfo{title}{Complex dynamics in nanosystems}.
\newblock \emph{\bibinfo{journal}{Phys. Rev. E}} \textbf{\bibinfo{volume}{87}},
  \bibinfo{pages}{052911} (\bibinfo{year}{2013}).

\bibitem{Alley:2003}
\bibinfo{author}{Alley, R.~B.} \emph{et~al.}
\newblock \bibinfo{title}{Abrupt climate change}.
\newblock \emph{\bibinfo{journal}{Science}} \textbf{\bibinfo{volume}{299}},
  \bibinfo{pages}{2005--2010} (\bibinfo{year}{2003}).

\bibitem{May:1977}
\bibinfo{author}{May, R.~M.}
\newblock \bibinfo{title}{Thresholds and breakpoints in ecosystems with a
  multiplicity of stable states}.
\newblock \emph{\bibinfo{journal}{Nature}} \textbf{\bibinfo{volume}{269}},
  \bibinfo{pages}{471--477} (\bibinfo{year}{1977}).

\bibitem{SPD:2005}
\bibinfo{author}{Schr{\"o}der, A.}, \bibinfo{author}{Persson, L.} \&
  \bibinfo{author}{De~Roos, A.~M.}
\newblock \bibinfo{title}{Direct experimental evidence for alternative stable
  states: a review}.
\newblock \emph{\bibinfo{journal}{Oikos}} \textbf{\bibinfo{volume}{110}},
  \bibinfo{pages}{3--19} (\bibinfo{year}{2005}).

\bibitem{Chase:2003}
\bibinfo{author}{Chase, J.~M.}
\newblock \bibinfo{title}{Experimental evidence for alternative stable
  equilibria in a benthic pond food web}.
\newblock \emph{\bibinfo{journal}{Ecol. Lett.}} \textbf{\bibinfo{volume}{6}},
  \bibinfo{pages}{733--741} (\bibinfo{year}{2003}).

\bibitem{BM:2005}
\bibinfo{author}{Badzey, R.~L.} \& \bibinfo{author}{Mohanty, P.}
\newblock \bibinfo{title}{Coherent signal amplification in bistable
  nanomechanical oscillators by stochastic resonance}.
\newblock \emph{\bibinfo{journal}{Nature}} \textbf{\bibinfo{volume}{437}},
  \bibinfo{pages}{995--998} (\bibinfo{year}{2005}).

\bibitem{WZXW:2011}
\bibinfo{author}{Wang, J.}, \bibinfo{author}{Zhang, K.}, \bibinfo{author}{Xu,
  L.} \& \bibinfo{author}{Wang, E.}
\newblock \bibinfo{title}{Quantifying the waddington landscape and biological
  paths for development and differentiation}.
\newblock \emph{\bibinfo{journal}{Proc. Natl. Acad. Sci. (USA)}}
  \textbf{\bibinfo{volume}{108}}, \bibinfo{pages}{8257--8262}
  (\bibinfo{year}{2011}).

\bibitem{HGME:2007}
\bibinfo{author}{Huang, S.}, \bibinfo{author}{Guo, Y.-P.},
  \bibinfo{author}{May, G.} \& \bibinfo{author}{Enver, T.}
\newblock \bibinfo{title}{Bifurcation dynamics in lineage-commitment in
  bipotent progenitor cells}.
\newblock \emph{\bibinfo{journal}{Developmental Biol.}}
  \textbf{\bibinfo{volume}{305}}, \bibinfo{pages}{695--713}
  (\bibinfo{year}{2007}).

\bibitem{Huang:2013}
\bibinfo{author}{Huang, S.}
\newblock \bibinfo{title}{Genetic and non-genetic instability in tumor
  progression: link between the fitness landscape and the epigenetic landscape
  of cancer cells}.
\newblock \emph{\bibinfo{journal}{Cancer Metastasis Rev.}}
  \textbf{\bibinfo{volume}{32}}, \bibinfo{pages}{423--448}
  (\bibinfo{year}{2013}).

\bibitem{SGLE:2006}
\bibinfo{author}{S{\"u}el, G.~M.}, \bibinfo{author}{Garcia-Ojalvo, J.},
  \bibinfo{author}{Liberman, L.~M.} \& \bibinfo{author}{Elowitz, M.~B.}
\newblock \bibinfo{title}{An excitable gene regulatory circuit induces
  transient cellular differentiation}.
\newblock \emph{\bibinfo{journal}{Nature}} \textbf{\bibinfo{volume}{440}},
  \bibinfo{pages}{545--550} (\bibinfo{year}{2006}).

\bibitem{HEYI:2005}
\bibinfo{author}{Huang, S.}, \bibinfo{author}{Eichler, G.},
  \bibinfo{author}{Bar-Yam, Y.} \& \bibinfo{author}{Ingber, D.~E.}
\newblock \bibinfo{title}{Cell fates as high-dimensional attractor states of a
  complex gene regulatory network}.
\newblock \emph{\bibinfo{journal}{Phys. Rev. Lett.}}
  \textbf{\bibinfo{volume}{94}}, \bibinfo{pages}{128701}
  (\bibinfo{year}{2005}).

\bibitem{FK:2012}
\bibinfo{author}{Furusawa, C.} \& \bibinfo{author}{Kaneko, K.}
\newblock \bibinfo{title}{A dynamical-systems view of stem cell biology}.
\newblock \emph{\bibinfo{journal}{Science}} \textbf{\bibinfo{volume}{338}},
  \bibinfo{pages}{215--217} (\bibinfo{year}{2012}).

\bibitem{LZW:2014}
\bibinfo{author}{Li, X.}, \bibinfo{author}{Zhang, K.} \& \bibinfo{author}{Wang,
  J.}
\newblock \bibinfo{title}{Exploring the mechanisms of differentiation,
  dedifferentiation, reprogramming and transdifferentiation}.
\newblock \emph{\bibinfo{journal}{PloS one}} \textbf{\bibinfo{volume}{9}},
  \bibinfo{pages}{e105216} (\bibinfo{year}{2014}).

\bibitem{YTWNJY:2011}
\bibinfo{author}{Yao, G.}, \bibinfo{author}{Tan, C.}, \bibinfo{author}{West,
  M.}, \bibinfo{author}{Nevins, J.} \& \bibinfo{author}{You, L.}
\newblock \bibinfo{title}{Origin of bistability underlying mammalian cell cycle
  entry}.
\newblock \emph{\bibinfo{journal}{Mol. Syst. Biol.}}
  \textbf{\bibinfo{volume}{7}} (\bibinfo{year}{2011}).

\bibitem{BT:2004}
\bibinfo{author}{Battogtokh, D.} \& \bibinfo{author}{Tyson, J.~J.}
\newblock \bibinfo{title}{Bifurcation analysis of a model of the budding yeast
  cell cycle}.
\newblock \emph{\bibinfo{journal}{Chaos}} \textbf{\bibinfo{volume}{14}},
  \bibinfo{pages}{653--661} (\bibinfo{year}{2004}).

\bibitem{KPST:2004}
\bibinfo{author}{Kauffman, S.}, \bibinfo{author}{Peterson, C.},
  \bibinfo{author}{Samuelsson, B.} \& \bibinfo{author}{Troein, C.}
\newblock \bibinfo{title}{Genetic networks with canalyzing boolean rules are
  always stable}.
\newblock \emph{\bibinfo{journal}{Proc. Natl. Acad. Sci. (USA)}}
  \textbf{\bibinfo{volume}{101}}, \bibinfo{pages}{17102--17107}
  (\bibinfo{year}{2004}).

\bibitem{GD:2005}
\bibinfo{author}{Greil, F.} \& \bibinfo{author}{Drossel, B.}
\newblock \bibinfo{title}{Dynamics of critical kauffman networks under
  asynchronous stochastic update}.
\newblock \emph{\bibinfo{journal}{Phys. Rev. Lett.}}
  \textbf{\bibinfo{volume}{95}}, \bibinfo{pages}{048701}
  (\bibinfo{year}{2005}).

\bibitem{MGAB:2008}
\bibinfo{author}{Motter, A.}, \bibinfo{author}{Gulbahce, N.},
  \bibinfo{author}{Almaas, E.} \& \bibinfo{author}{Barab{\'a}si, A.-L.}
\newblock \bibinfo{title}{Predicting synthetic rescues in metabolic networks}.
\newblock \emph{\bibinfo{journal}{Mol. Sys. Biol.}}
  \textbf{\bibinfo{volume}{4}} (\bibinfo{year}{2008}).

\bibitem{MTELT:2009}
\bibinfo{author}{Ma, W.}, \bibinfo{author}{Trusina, A.},
  \bibinfo{author}{El-Samad, H.}, \bibinfo{author}{Lim, W.~A.} \&
  \bibinfo{author}{Tang, C.}
\newblock \bibinfo{title}{Defining network topologies that can achieve
  biochemical adaptation}.
\newblock \emph{\bibinfo{journal}{Cell}} \textbf{\bibinfo{volume}{138}},
  \bibinfo{pages}{760--773} (\bibinfo{year}{2009}).

\bibitem{Faucon:2014}
\bibinfo{author}{Faucon, P.~C.} \emph{et~al.}
\newblock \bibinfo{title}{Gene networks of fully connected triads with complete
  auto-activation enable multistability and stepwise stochastic transitions}.
\newblock \emph{\bibinfo{journal}{PloS one}} \textbf{\bibinfo{volume}{9}},
  \bibinfo{pages}{e102873} (\bibinfo{year}{2014}).

\bibitem{Gardner:2000}
\bibinfo{author}{Gardner, T.~S.}, \bibinfo{author}{Cantor, C.~R.} \&
  \bibinfo{author}{Collins, J.~J.}
\newblock \bibinfo{title}{Construction of a genetic toggle switch in
  escherichia coli}.
\newblock \emph{\bibinfo{journal}{Nature}} \textbf{\bibinfo{volume}{403}},
  \bibinfo{pages}{339--342} (\bibinfo{year}{2000}).

\bibitem{WSLELW:2013}
\bibinfo{author}{Wu, M.} \emph{et~al.}
\newblock \bibinfo{title}{Engineering of regulated stochastic cell fate
  determination}.
\newblock \emph{\bibinfo{journal}{Proc. Natl. Acad. Sci. (USA)}}
  \textbf{\bibinfo{volume}{110}}, \bibinfo{pages}{10610--10615}
  (\bibinfo{year}{2013}).

\bibitem{WMX:2014}
\bibinfo{author}{Wu, F.}, \bibinfo{author}{Menn, D.} \& \bibinfo{author}{Wang,
  X.}
\newblock \bibinfo{title}{Quorum-sensing crosstalk-driven synthetic circuits:
  From unimodality to trimodality}.
\newblock \emph{\bibinfo{journal}{Chem. Biol.}} \textbf{\bibinfo{volume}{21}},
  \bibinfo{pages}{1629--1638} (\bibinfo{year}{2014}).

\bibitem{Lai:2014}
\bibinfo{author}{Lai, Y.-C.}
\newblock \bibinfo{title}{Controlling complex, nonlinear dynamical networks}.
\newblock \emph{\bibinfo{journal}{Nat. Sci. Rev.}}
  \textbf{\bibinfo{volume}{1}}, \bibinfo{pages}{339--341}
  (\bibinfo{year}{2014}).

\bibitem{Feala:2010}
\bibinfo{author}{Feala, J.~D.} \emph{et~al.}
\newblock \bibinfo{title}{Systems approaches and algorithms for discovery of
  combinatorial therapies}.
\newblock \emph{\bibinfo{journal}{Wiley Interdiscip. Rev. Syst. Biol. Med.}}
  \textbf{\bibinfo{volume}{2}}, \bibinfo{pages}{181--193}
  (\bibinfo{year}{2010}).

\bibitem{Fitzgerald:2006}
\bibinfo{author}{Fitzgerald, J.~B.}, \bibinfo{author}{Schoeberl, B.},
  \bibinfo{author}{Nielsen, U.~B.} \& \bibinfo{author}{Sorger, P.~K.}
\newblock \bibinfo{title}{Systems biology and combination therapy in the quest
  for clinical efficacy}.
\newblock \emph{\bibinfo{journal}{Nat. Chem. Biol.}}
  \textbf{\bibinfo{volume}{2}}, \bibinfo{pages}{458--466}
  (\bibinfo{year}{2006}).

\bibitem{Zhang:2008}
\bibinfo{author}{Zhang, R.} \emph{et~al.}
\newblock \bibinfo{title}{Network model of survival signaling in large granular
  lymphocyte leukemia}.
\newblock \emph{\bibinfo{journal}{Proc. Natl. Acad. Sci. (USA)}}
  \textbf{\bibinfo{volume}{105}}, \bibinfo{pages}{16308--16313}
  (\bibinfo{year}{2008}).

\bibitem{Saadatpour:2011}
\bibinfo{author}{Saadatpour, A.} \emph{et~al.}
\newblock \bibinfo{title}{Dynamical and structural analysis of a t cell
  survival network identifies novel candidate therapeutic targets for large
  granular lymphocyte leukemia}.
\newblock \emph{\bibinfo{journal}{PloS Comp. Biol.}}
  \textbf{\bibinfo{volume}{7}}, \bibinfo{pages}{e1002267}
  (\bibinfo{year}{2011}).

\bibitem{Wittmann:2009}
\bibinfo{author}{Wittmann, D.~M.} \emph{et~al.}
\newblock \bibinfo{title}{Transforming boolean models to continuous models:
  methodology and application to t-cell receptor signaling}.
\newblock \emph{\bibinfo{journal}{BMC Syst. Biol.}}
  \textbf{\bibinfo{volume}{3}}, \bibinfo{pages}{98} (\bibinfo{year}{2009}).

\bibitem{KPWT:2010}
\bibinfo{author}{Krumsiek, J.}, \bibinfo{author}{P\"{o}lsterl, S.},
  \bibinfo{author}{Wittmann, D.~M.} \& \bibinfo{author}{Theis, F.~J.}
\newblock \bibinfo{title}{Odefy-from discrete to continuous models}.
\newblock \emph{\bibinfo{journal}{BMC Bioinformatcis}}
  \textbf{\bibinfo{volume}{11}}, \bibinfo{pages}{233} (\bibinfo{year}{2010}).

\bibitem{CKM:2013}
\bibinfo{author}{Cornelius, S.~P.}, \bibinfo{author}{Kath, W.~L.} \&
  \bibinfo{author}{Motter, A.~E.}
\newblock \bibinfo{title}{Realistic control of network dynamics}.
\newblock \emph{\bibinfo{journal}{Nat. Comm.}} \textbf{\bibinfo{volume}{4}},
  \bibinfo{pages}{1942} (\bibinfo{year}{2013}).

\bibitem{BSV:1981}
\bibinfo{author}{Benzi, R.}, \bibinfo{author}{Sutera, A.} \&
  \bibinfo{author}{Vulpiani, A.}
\newblock \bibinfo{title}{The mechanism of stochastic resonance}.
\newblock \emph{\bibinfo{journal}{J. Phys. A}} \textbf{\bibinfo{volume}{14}},
  \bibinfo{pages}{L453--L457} (\bibinfo{year}{1981}).

\bibitem{BPSV:1983}
\bibinfo{author}{Benzi, R.}, \bibinfo{author}{Parisi, G.},
  \bibinfo{author}{Sutera, A.} \& \bibinfo{author}{Vulpiani, A.}
\newblock \bibinfo{title}{A theory of stochastic resonance in climatic-change}.
\newblock \emph{\bibinfo{journal}{J. Appl. Math.}}
  \textbf{\bibinfo{volume}{43}}, \bibinfo{pages}{565--578}
  (\bibinfo{year}{1983}).

\bibitem{MW:1989}
\bibinfo{author}{McNamara, B.} \& \bibinfo{author}{Wiesenfeld, K.}
\newblock \bibinfo{title}{Theory of stochastic resonance}.
\newblock \emph{\bibinfo{journal}{Phys. Rev. A}} \textbf{\bibinfo{volume}{39}},
  \bibinfo{pages}{4854--4869} (\bibinfo{year}{1989}).

\bibitem{MPO:1994}
\bibinfo{author}{Moss, F.}, \bibinfo{author}{Pierson, D.} \&
  \bibinfo{author}{O'Gorman, D.}
\newblock \bibinfo{title}{Stochastic resonance - tutorial and update}.
\newblock \emph{\bibinfo{journal}{Int. J. Bif. Chaos}}
  \textbf{\bibinfo{volume}{4}}, \bibinfo{pages}{1383--1397}
  (\bibinfo{year}{1994}).

\bibitem{GNCM:1997}
\bibinfo{author}{Gailey, P.~C.}, \bibinfo{author}{Neiman, A.},
  \bibinfo{author}{Collins, J.~J.} \& \bibinfo{author}{Moss, F.}
\newblock \bibinfo{title}{Stochastic resonance in ensembles of nondynamical
  elements: The role of internal noise}.
\newblock \emph{\bibinfo{journal}{Phys. Rev. Lett.}}
  \textbf{\bibinfo{volume}{79}}, \bibinfo{pages}{4701--4704}
  (\bibinfo{year}{1997}).

\bibitem{GHJM:1998}
\bibinfo{author}{Gammaitoni, L.}, \bibinfo{author}{H\"{a}nggi, P.},
  \bibinfo{author}{Jung, P.} \& \bibinfo{author}{Marchesoni, F.}
\newblock \bibinfo{title}{Stochastic resonance}.
\newblock \emph{\bibinfo{journal}{Rev. Mod. Phys.}}
  \textbf{\bibinfo{volume}{70}}, \bibinfo{pages}{223--287}
  (\bibinfo{year}{1998}).

\bibitem{SH:1989}
\bibinfo{author}{Sigeti, D.} \& \bibinfo{author}{Horsthemke, W.}
\newblock \bibinfo{title}{Pseudo-regular oscillations induced by external
  noise}.
\newblock \emph{\bibinfo{journal}{J. Stat. Phys.}}
  \textbf{\bibinfo{volume}{54}}, \bibinfo{pages}{1217} (\bibinfo{year}{1989}).

\bibitem{PK:1997}
\bibinfo{author}{Pikovsky, A.~S.} \& \bibinfo{author}{Kurths, J.}
\newblock \bibinfo{title}{Coherence resonance in a noise-driven excitable
  system}.
\newblock \emph{\bibinfo{journal}{Phys. Rev. Lett.}}
  \textbf{\bibinfo{volume}{78}}, \bibinfo{pages}{775--778}
  (\bibinfo{year}{1997}).

\bibitem{LL:2001a}
\bibinfo{author}{Liu, Z.} \& \bibinfo{author}{Lai, Y.-C.}
\newblock \bibinfo{title}{Coherence resonance in coupled chaotic oscillators}.
\newblock \emph{\bibinfo{journal}{Phys. Rev. Lett.}}
  \textbf{\bibinfo{volume}{86}}, \bibinfo{pages}{4737--4740}
  (\bibinfo{year}{2001}).

\bibitem{LL:2001b}
\bibinfo{author}{Lai, Y.-C.} \& \bibinfo{author}{Liu, Z.}
\newblock \bibinfo{title}{Noise-enhanced temporal regularity in coupled chaotic
  oscillators}.
\newblock \emph{\bibinfo{journal}{Phys. Rev. E}} \textbf{\bibinfo{volume}{64}},
  \bibinfo{pages}{066202} (\bibinfo{year}{2001}).

\bibitem{Yin:2014}
\bibinfo{author}{Yin, N.} \emph{et~al.}
\newblock \bibinfo{title}{Synergistic and antagonistic drug combinations depend
  on network topology}.
\newblock \emph{\bibinfo{journal}{PloS one}} \textbf{\bibinfo{volume}{9}},
  \bibinfo{pages}{e93960} (\bibinfo{year}{2014}).

\bibitem{BZA:2011}
\bibinfo{author}{Bhattacharya, S.}, \bibinfo{author}{Zhang, Q.} \&
  \bibinfo{author}{Andersen, M.~E.}
\newblock \bibinfo{title}{A deterministic map of waddington's epigenetic
  landscape for cell fate specification}.
\newblock \emph{\bibinfo{journal}{BMC Syst. Biol.}}
  \textbf{\bibinfo{volume}{5}}, \bibinfo{pages}{85} (\bibinfo{year}{2011}).

\bibitem{KST:2013}
\bibinfo{author}{Kromer, J.~A.}, \bibinfo{author}{Schimansky-Geier, L.} \&
  \bibinfo{author}{Toral, R.}
\newblock \bibinfo{title}{Weighted-ensemble brownian dynamics simulation:
  Sampling of rare events in nonequilibrium systems}.
\newblock \emph{\bibinfo{journal}{Phys. Rev. E}} \textbf{\bibinfo{volume}{87}},
  \bibinfo{pages}{063311} (\bibinfo{year}{2013}).

\bibitem{Waddington:book}
\bibinfo{author}{Waddington, C.~H.}
\newblock \emph{\bibinfo{title}{The Strategy of the Genes}}
  (\bibinfo{publisher}{Allen $\&$ Unwin}, \bibinfo{address}{London},
  \bibinfo{year}{1957}).

\bibitem{Huang:2009}
\bibinfo{author}{Huang, S.}
\newblock \bibinfo{title}{Reprogramming cell fates: reconciling rarity with
  robustness}.
\newblock \emph{\bibinfo{journal}{Bioessays}} \textbf{\bibinfo{volume}{31}},
  \bibinfo{pages}{546--560} (\bibinfo{year}{2009}).

\bibitem{MML:2009}
\bibinfo{author}{MacArthur, B.~D.}, \bibinfo{author}{Maayan, A.} \&
  \bibinfo{author}{Lemischka, I.~R.}
\newblock \bibinfo{title}{Systems biology of stem cell fate and cellular
  reprogramming}.
\newblock \emph{\bibinfo{journal}{Nat. Rev. Mol. Cell Biol.}}
  \textbf{\bibinfo{volume}{10}}, \bibinfo{pages}{672--681}
  (\bibinfo{year}{2009}).

\bibitem{CS:2012}
\bibinfo{author}{Corson, F.} \& \bibinfo{author}{Siggia, E.~D.}
\newblock \bibinfo{title}{Geometry, epistasis, and developmental patterning}.
\newblock \emph{\bibinfo{journal}{Proc. Nat. Acad. Sci. (USA)}}
  \textbf{\bibinfo{volume}{109}}, \bibinfo{pages}{5568--5575}
  (\bibinfo{year}{2012}).

\bibitem{WXW:2008}
\bibinfo{author}{Wang, J.}, \bibinfo{author}{Xu, L.} \& \bibinfo{author}{Wang,
  E.~K.}
\newblock \bibinfo{title}{Potential landscape and flux framework of
  nonequilibrium networks: Robustness, dissipation, and coherence of
  biochemical oscillations}.
\newblock \emph{\bibinfo{journal}{Proc. Natl. Acad. Sci. USA}}
  \textbf{\bibinfo{volume}{105}}, \bibinfo{pages}{12271--12276}
  (\bibinfo{year}{2008}).

\bibitem{WXWH:2010}
\bibinfo{author}{Wang, J.}, \bibinfo{author}{Xu, L.}, \bibinfo{author}{Wang,
  E.~K.} \& \bibinfo{author}{Huang, S.}
\newblock \bibinfo{title}{The potential landscape of genetic circuits imposes
  the arrow of time in stem cell differentiation}.
\newblock \emph{\bibinfo{journal}{Biophys. J.}} \textbf{\bibinfo{volume}{99}},
  \bibinfo{pages}{29--39} (\bibinfo{year}{2010}).

\bibitem{ZXZWW:2012}
\bibinfo{author}{Zhang, F.}, \bibinfo{author}{Xu, L.}, \bibinfo{author}{Zhang,
  K.}, \bibinfo{author}{Wang, E.~K.} \& \bibinfo{author}{Wang, J.}
\newblock \bibinfo{title}{The potential and flux landscape theory of
  evolution}.
\newblock \emph{\bibinfo{journal}{J. Chem. Phys.}}
  \textbf{\bibinfo{volume}{137}}, \bibinfo{pages}{065102}
  (\bibinfo{year}{2012}).

\bibitem{GT:1984}
\bibinfo{author}{Garham, R.} \& \bibinfo{author}{T\'el, T.}
\newblock \bibinfo{title}{Existence of a potential for dissipative dynamical
  systems}.
\newblock \emph{\bibinfo{journal}{Phys. Rev. Lett.}}
  \textbf{\bibinfo{volume}{52}}, \bibinfo{pages}{9--12} (\bibinfo{year}{1984}).

\bibitem{GHT:1991}
\bibinfo{author}{Graham, R.}, \bibinfo{author}{Hamm, A.} \&
  \bibinfo{author}{T\'el, T.}
\newblock \bibinfo{title}{Nonequilibrium potentials for dynamical systems with
  fractal attractors and repellers}.
\newblock \emph{\bibinfo{journal}{Phys. Rev. Lett.}}
  \textbf{\bibinfo{volume}{66}}, \bibinfo{pages}{3089--3092}
  (\bibinfo{year}{1991}).

\bibitem{TL:2010}
\bibinfo{author}{T\'el, T.} \& \bibinfo{author}{Lai, Y.-C.}
\newblock \bibinfo{title}{Quasipotential approach to critical scaling in
  noise-induced chaos}.
\newblock \emph{\bibinfo{journal}{Phys. Rev. E}} \textbf{\bibinfo{volume}{81}},
  \bibinfo{pages}{056208} (\bibinfo{year}{2010}).

\bibitem{Shuetal:2013}
\bibinfo{author}{Shu, J.} \emph{et~al.}
\newblock \bibinfo{title}{Induction of pluripotency in mouse somatic cells with
  lineage specifiers}.
\newblock \emph{\bibinfo{journal}{Cell}} \textbf{\bibinfo{volume}{153}},
  \bibinfo{pages}{963--975} (\bibinfo{year}{2013}).

\bibitem{OGY:1990}
\bibinfo{author}{Ott, E.}, \bibinfo{author}{Grebogi, C.} \&
  \bibinfo{author}{Yorke, J.~A.}
\newblock \bibinfo{title}{Controlling chaos}.
\newblock \emph{\bibinfo{journal}{Phys. Rev. Lett.}}
  \textbf{\bibinfo{volume}{64}}, \bibinfo{pages}{1196--1199}
  (\bibinfo{year}{1990}).

\bibitem{BGLMM:2000}
\bibinfo{author}{Boccaletti, S.}, \bibinfo{author}{Grebogi, C.},
  \bibinfo{author}{Lai, Y.-C.}, \bibinfo{author}{Mancini, H.} \&
  \bibinfo{author}{Maza, D.}
\newblock \bibinfo{title}{Control of chaos: theory and applications}.
\newblock \emph{\bibinfo{journal}{Phys. Rep.}} \textbf{\bibinfo{volume}{329}},
  \bibinfo{pages}{103--197} (\bibinfo{year}{2000}).

\bibitem{SL:book}
\bibinfo{author}{Slotine, J.-J.~E.} \& \bibinfo{author}{Li, W.}
\newblock \emph{\bibinfo{title}{Applied Nonlinear Control}}
  (\bibinfo{publisher}{Prentice-Hall}, \bibinfo{address}{New Jersey},
  \bibinfo{year}{1991}).

\bibitem{Ott:book}
\bibinfo{author}{Ott, E.}
\newblock \emph{\bibinfo{title}{Chaos in Dynamical Systems}}
  (\bibinfo{publisher}{Cambridge University Press},
  \bibinfo{address}{Cambridge, UK}, \bibinfo{year}{2002}),
  \bibinfo{edition}{second} edn.

\end{thebibliography}

\section*{Acknowledgement}

This work was supported by ARO under Grant W911NF-14-1-0504.

\section*{Author contributions}

Devised the research project: YCL, XW, and CG; 
Performed numerical simulations: LZW and RS; 
Analyzed the results: LZW, RS, ZGH, YCL, and WXW;
Wrote the paper: YCL, LZW, ZGH, and RS.

\section*{Additional information}

{\bf Supplementary Information} accompanies this paper.

\section*{Competing financial interests}

The authors declare no competing financial interests.
\end{document}